\documentclass[11pt]{article}
\usepackage{amsmath,amsthm}
\usepackage{latexsym}
\usepackage{amssymb}
\usepackage{array}
\usepackage{epsfig}
\newtheorem{teo}{Theorem}
\newtheorem{lem}[teo]{Lemma}
\newtheorem{cor}[teo]{Corollary}

\newtheorem{con}{Conjecture}

\newcommand{\Epsfig}[5] {
                          \begin{figure}[#2]
                              \epsfxsize=#5
                              \centerline{\epsfbox{#1}}
                              \caption{\small{#3} \label{#4}}
                          \end{figure}
                        }
\newcommand{\EpsfigR}[5] {
                          \begin{figure}[#2]
                              \centerline{\epsfig{file=#1,angle=-90,width=#5}}
                              \caption{\small{#3} \label{#4}}
                          \end{figure}
                        }
\newcommand{\Sec}[2]    { \section{#1}   \label{#2}}         

\newcommand{\eq}[1]     {Eq.~(\ref{#1})}
\newcommand{\Th}[1]     {Theorem~\ref{#1}}
\newcommand{\Lem}[1]    {Lemma~\ref{#1}}

\newcommand{\Fig}[1]    {Fig.~\ref{#1}}
\newcommand{\Secref}[1] {\S\ref{#1}}
\newcommand{\Tblref}[1] {Table~\ref{#1}}

\newcommand{\makev}[2]{\left( 
                         \begin{array}{c}{#1}\\{#2}\end{array}\right)}

\newcommand{\hen}{H\'enon }
\newcommand{\tblrule}{\rule[-1.2mm]{0mm}{4mm}}
\newcommand{\per}[1]{(#1)^{\infty}}
\newcommand{\homoc}[1]{+^\infty-(#1)-+^\infty}
\newcommand{\homo}[1]{(#1)}

\newcommand{\ksn}{k_{\rm sn}}
\newcommand{\kpf}{k_{\rm pf}}
\newcommand{\esn}{\epsilon_{\rm sn}}
\newcommand{\asn} {\mbox{\rm asn}} 
\newcommand{\sn} {\mbox{\rm sn}} 
\newcommand{\sigmaF} {\Sigma_{\cal{F}}}
\newcommand{\pf} {\mbox{\rm pf}} 
\newcommand{\pd}{\mbox{\rm pd}}

\newcommand{\bif}[2]{#1 \, \{ #2 \}  }
\newcommand{\ptcbif}[3]{#1 \rightarrow \bif{#2}{#3}  } 

\renewcommand{\qed} {$\Box$ \\}
\newcommand{\qedd} {$\Box$ }
\newcommand{\Ws} {{W}^{s}}
\newcommand{\Wu} {{W}^{u}}
\newcommand{\sdot}{\mbox{{\Huge .}}}

\newcommand{\Tr}{\mathop{\rm Tr}}
\newcommand{\sign}{\mathop{\rm sign}}
\newcommand{\type}{\mathop{\rm type}}
\newcommand{\fix}{\mathop{\rm fix}}
\newcommand{\trans}{{\rm trans}}
\newcommand{\bG}{\mathbf{G}}

\newcommand{\bs} {\mathbf{s}}

\newcommand{\bT} {\mathbf{T}}
\newcommand{\bH} {\mathbf{H}}
\newcommand{\bz} {\mathbf{z}}
\newcommand{\diag}{\mbox{diag}}
\newcommand{\id} {\mbox{id}}

\newcommand{\mR} {\mathbb{R}}
\newcommand{\mZ} {\mathbb{Z}}

\newcommand{\balpha} {\beta}

\begin{document}

\title{Homoclinic Bifurcations for the \hen Map}
\author{D. Sterling,  H.R. Dullin \& J.D. Meiss 
\thanks{Useful conversations with R. Easton and B. Peckham 
are gratefully acknowledged.  DS was supported in part by NSF 
traineeship grant number DMS-9208685, JDM was supported in part by NSF 
grant number DMS-9623216 and HRD was supported by DFG grant number Du 302}\\
Department of Applied Mathematics\\
University of Colorado\\
Boulder, CO 80309
}
\maketitle
\begin{abstract} Chaotic dynamics can be effectively studied by 
continuation from an anti-integrable limit.  We use this limit to 
assign global symbols to orbits and use continuation from the limit to 
study their bifurcations.  We find a bound on the parameter range for 
which the \hen map exhibits a complete binary horseshoe as well as a 
subshift of finite type.
We classify  homoclinic bifurcations, and study those for the 
area preserving case in detail. 
Simple forcing relations between homoclinic orbits are established.
We show that a symmetry of the map gives rise to constraints on
certain sequences of homoclinic bifurcations.
Our numerical studies also identify the bifurcations that bound
intervals on which the topological entropy is apparently constant.
\\
\textbf{AMS classification scheme numbers}: 58F05, 58F03, 58C15
%
%
\end{abstract}

%
\Sec{Introduction}{introsec}

The problem of determining the sequence of bifurcations that result in 
the creation of a Smale horseshoe is an interesting one 
\cite{Easton86, Grassberger89, Davis91}.  In this paper we use a 
continuation technique based on an ``anti-integrable'' (AI) limit 
\cite{Aubry91} to study some of these bifurcations from the opposite 
side, that is, as bifurcations that destroy the horseshoe.

As a simple example, we study the family of \hen maps
\cite{Henon68,Henon76}
\begin{equation} \label{henmap}
     \makev {{x'}}{{y'}} = \makev{ {y-k+x^2} }{ {-bx}} .
\end{equation}
Apart from the fact that the \hen maps are the simplest, non-trivial maps 
of the plane, they are of more general interest as well, since vector 
fields in the neighborhood of certain codimension-two homoclinic 
bifurcations can be reduced to H\'enon-like maps 
\cite{Naudot96,CHS96}.

As we recall in \Secref{sec:ai}, the AI limit for this map is 
essentially $k \rightarrow \infty$.  In order to represent this with 
finite parameters, we need only rescale the map, letting 
\[
	z = \epsilon  x \;, \mbox{ where } \;
	\epsilon  = \frac{1}{\sqrt{k}} \; .
\]

As was shown by Devaney and Nitecki \cite{Devaney79}, the \hen map 
has a hyperbolic horseshoe when
\begin{equation} \label{krange}
        k >  (1+|b|)^{2} \frac{5+2\sqrt5}{4} \; .
\end{equation}
The \hen map has at most $2^{n}$ \cite{Moser60}
periodic points of period $n$, and when 
the map has a hyperbolic horseshoe, all these orbits exist and are 
easily identified by their symbolic labels.

We showed earlier \cite{Sterling98a} that a contraction mapping 
argument implies there is a one-to-one correspondence between orbits 
in the AI limit and bounded orbits of the \hen map in precisely the 
range \eq{krange}.  In \Secref{sec:subshift} we show that if we consider a 
particular subset of orbits, this bound can be increased.  
Moreover, in \Secref{sec:horse}, we present results of numerical 
investigations for all $b$ that give what we believe are optimal bounds.

In general, the existence of an anti-integrable limit leads to a 
natural symbolic characterization of orbits---for the \hen map this is 
the same as the horseshoe coding.  We use this coding, and as 
discussed in \Secref {sec:numerics}, a predictor-corrector 
continuation method \cite{Seydel}, to give each orbit a {\it global 
code}.  
That is, we label an orbit with the AI code, and use this designation 
for the family of orbits until it collides with a family with a 
different code.  For the \hen map, this gives a map from the bounded 
orbits of the map to sequences of symbols $\bs \in 2^{\mZ}$ modulo 
cyclic permutations, providing only that every orbit can be {\em 
smoothly} connected to the AI limit.
This is the working hypothesis for our numerical method, even though
we know that it is probably not true in general. 
It is certainly valid when the hyperbolic horseshoe exists.
We give an example of a periodic orbit not smoothly connected
to the AI limit in the dissipative case. This illustrates a
general anti-monotonicity result \cite{Yorke92}, 
stating  that when $b \not = -1,0,1$ the map generically 
creates but also destroys orbits when $k$ is increased, so that
the topological entropy is not necessarily monotone.
We extend this anti-monotonicity result to the area preserving
case by a quite different argument concerning the vanishing of
twist in the neighborhood of the period tripling bifurcations 
in a separate paper \cite{DMS99}.
Even though the area preserving case exhibits anti-monotonicity,
we still conjecture that there are no isolated bubbles in the bifurcation
diagram, i.e.\ that every orbit is continuously connected
to the AI limit.

Our global code contrasts with other methods for constructing symbolic 
dynamics for maps, which rely on some attempt to obtain a generating 
partition \cite{Cvitanovic88, Grassberger89, Hao91, 
Hansen92, Christiansen95b}.  These methods rely on somewhat 
ad hoc techniques for constructing the partition, especially when 
there exist elliptic orbits.  Our method gives a natural partition that 
is smoothly connected to the horseshoe, though it does rely on our
working hypothesis

In our computations of the \hen map, we observe that the horseshoe 
destroying bifurcation appears to be a homoclinic saddle-node 
bifurcation when the map is orientation preserving, and a heteroclinic 
saddle-node when it is not.  In \Secref{sec:homo} we study in detail 
the homoclinic orbits of the area preserving \hen map, and show how 
the AI code directly gives other properties of the orbits, such as 
their ``type,'' ``transition time,'' and ``Poincar\'e signature.''

For an area preserving map, the destruction of a horseshoe by a 
homoclinic bifurcation gives rise generically to elliptic orbits.  
Specifically, if $f$ is a $ C^{1}$ area preserving diffeomorphism with 
a homoclinic tangency at $x$ then for any neighborhood $U$ of $x$, 
there is an area preserving diffeomorphism $C^{1}$-close to $f$ that
has an elliptic periodic point in $U$ \cite{New77}.

Much more is known about the behavior of area-contracting maps near a 
homoclinic tangency.  Gavrilov and Silnikov proved that if a $C^{3}$ 
map has a quadratic homoclinic tangency at a parameter $k^{*}$ then there 
exists a sequence of parameter values $k_{n} \rightarrow k^{*}$ such 
that at $k_{n}$ there is a saddle-node bifurcation creating orbits of 
period $n$ \cite{Gav72,Gav73}; because one of the created orbits is a 
sink, this is called a {\it cascade of sinks}.  Robinson extended 
these results to the real analytic case where the intersection is 
created degenerately \cite{Rob83}. In our computations we will find a 
similar cascade of saddle-node bifurcations for the area preserving 
\hen map---this gives a sequence of elliptic orbits that limit on the 
homoclinic bifurcation. Thus the destruction of the horseshoe is 
associated with the creation of the first stable ``island.''

The ordering on the invariant manifolds poses severe restrictions
on the possible bifurcations. In \Secref{sec:hobs} we use it
to prove which homoclinic bifurcation of the hyperbolic
fixed point is the first one. We observe that the forcing
relations between homoclinic orbits up to type 6 is essentially
like the unimodal ordering of one dimensional maps. 
Generically a homoclinic bifurcation corresponds to a quadratic 
tangency of the stable and unstable manifolds---a ``homoclinic 
saddle-node bifurcation.'' 
There are two more generic bifurcations in maps with a symmetry: a 
homoclinic pitchfork when the manifolds exhibit a cubic tangency, and 
a simultaneous pair of asymmetric saddle node bifurcations.  
In \Secref{sec:pitchfork} we show that a symmetric homoclinic 
bifurcation forbids certain other bifurcations to occur after it, 
leading to a natural mechanism to create homoclinic pitchfork 
bifurcations or asymmetric saddle node pairs. 
We observe all three of these bifurcations for the area preserving \hen 
map, which has a time-reversal symmetry.

Davis, MacKay, and Sannami \cite{Davis91} conjectured that there are 
intervals of $k$ below the horseshoe for which the \hen map is a 
hyperbolic Markov shift.  They also identified the Markov partitions 
for these cases.  Their conjecture was based on computing all the 
periodic orbits up to a period $20$ using the technique of Biham and 
Wenzel \cite{Biham89, Biham90}.  In \Secref{sec:entropy}, we confirm 
their computations with our continuation technique and extend them to period 
$24$---an order of magnitude more orbits. Moreover, we identify 
the bifurcations responsible for the creation and destruction of these 
apparently hyperbolic intervals; as befits with the theme of this 
paper, they are homoclinic bifurcations.


\Sec{Anti-Integrable Limit}{sec:ai}

Dynamics in discrete time can be represented by a relation, $F(x,x') = 
0$ where $x$ and $x'$ are points in some manifold.  Normally, 
we can explicitly solve for $x' = f(x)$, giving a map on the manifold, 
with orbits defined by sequences $x_{t}=f(x_{t-1})$.  Suppose, 
however, that $F$ depends upon a parameter $\epsilon$, in such a way 
that this is not always possible; for example, $F(x,x') = \epsilon 
G(x,x') + H(x)$.  In this case the implicit equation $F=0$ can no 
longer be solved for $x'$ when $\epsilon = 0$; instead ``orbits'' 
correspond to arbitrary sequences of points, $x_{t}$ that are zeros of 
$H$---the dynamics are not deterministic.  In this case we say that 
$\epsilon=0$ corresponds to an {\it anti-integrable (AI)} limit 
of the map $f$ \cite{Aubry95}.  If the derivative of $H$ is nonsingular, then 
a straightforward implicit function argument can be used to show that 
the AI orbits can be continued for $\epsilon \neq 0$ 
to orbits of the map $f$ \cite{Aubry91, MacKay92}.  An AI limit 
with this property is called {\it nondegenerate}.

For example, consider the \hen map \eq{henmap}.  Denoting points on an 
orbit by a sequence $x_{t}$, $t \in \mZ$, we can rewrite
\eq{henmap} as a second order difference equation
\[
     x_{t+1} + bx_{t-1} + k -x_t^2 = 0 \;. 
\]
Introducing the scaled coordinate $z = \epsilon x$ and choosing 
$\epsilon = k^{-1/2}$ gives an implicit map in the variable $z$ 
with parameter $\epsilon$
\begin{equation} \label{impmap}
     \epsilon(z_{t+1} + bz_{t-1}) + 1 -z_t^2 = 0 \;.
\end{equation}
With this choice of $\epsilon$, we can study only the range $ 0 < k < 
\infty$; however, one could redefine $\epsilon$ to shift this range.\footnote{
	For example, choosing $\hat\epsilon = (k+\delta)^{-1/2}$, maps 
	positive values of $\hat\epsilon$ to the range $-\delta < k < \infty$.  
	Our numerical routines typically use $\delta =1$ so that we can cover 
	the entire parameter range where there are bounded orbits.
	In the text we always use $\delta = 0$.
}
A period $n$ orbit of the \hen map is given by a sequence 
$z_0,z_1,\ldots,z_{n-1}$ that satisfies \eq{impmap}, together with the 
condition that $z_{t+n}=z_{t}$. The corresponding family of periodic orbits 
is denoted by $\bz(\epsilon)$.

At the AI limit, the map \eq{impmap} reduces to
\[ 
   z_t^2 = 1 \;.
\]
Thus ``orbits'' in this limit are arbitrary sequences of $\pm1$, which 
we abbreviate with $+$ and $-$.  Let $\Sigma$ denote the space of such 
sequences
\begin{equation}
\Sigma \equiv \{-,+\}^{\mZ} = \{ \bs: s_{t}\in\{-,+\} \,, t \in \mZ \} \,.
\end{equation}

For ease of notation we denote the corresponding sequence of 
$\{+1,-1\} \in \mR^\mZ$ by the same symbol $\bs$.
Hence every sequence $\bs \in \Sigma$ is an orbit $\bs \in \mR^\mZ$ at the 
anti-integrable limit, and each of these can be continued to an orbit 
of the \hen map for small enough $\epsilon$ \cite{Aubry90,MacKay92}.  
Previously we gave an explicit upper bound on $\epsilon$ for the 
existence of orbits for every symbol sequence \cite{Sterling98a}:

\begin{teo} \label{thrm:exist}
        For every symbol sequence $\bs \in \Sigma$ there exists a unique orbit 
        $\bz(\epsilon)$ of the \hen map \eq{impmap}, such that $\bz(0) = \bs$ 
        providing
        \begin{equation}\label{epsmax}
             |\epsilon|(1+|b|) < 2\sqrt{1-2/\sqrt{5}} \approx 0.649839  \,.
        \end{equation} 
\end{teo}

The basic idea of the proof of this theorem is as follows 
\cite{Sterling98a}.  Let $B_{M}$ be the closed ball of radius $M$ 
around the point $\bs \in \Sigma$,
\begin{equation}\label{BMdef}
    B_{M}(\bs) = \{{\bz}  : ||\bz - \bs||_{\infty} \le M \} \;.
\end{equation}
For each symbol sequence $\bs \in \Sigma$ and small enough $M$, define 
a map $\bT:B_M \rightarrow B_M$ by
\begin{equation}\label{Tdef}
  T_i(\bz) \equiv s_i\sqrt{1 + \epsilon(z_{i+1} + bz_{i-1})} \;,
\end{equation}
then the corresponding \hen map orbit $\bz(\epsilon)$ is a fixed point 
of $\bT$.  The conclusion of \Th{thrm:exist} follows from finding the 
maximum value of $\epsilon$ for which there is an $n$ such that 
$\bT^n$ is a contraction mapping (i.e., $\bT^n:B_M \rightarrow B_M 
\quad \mbox{and} \quad ||D\bT^n|| < 1$).

The fact that $\bT$ is a contraction implies that there are no
bifurcations in the range \eq{epsmax}. This is the statement of 
\begin{cor}
	There are no bifurcations in the \hen map when $\epsilon$ and $b$ are
	in the range given in \Th{thrm:exist}.
\end{cor}
{\bf Proof:} 
Denote the system of equations (\ref{impmap}) by $\bH(\bz,\epsilon) = 0$. 
This infinite continuation problem has a unique solution 
$z(\epsilon)$ if the inverse of $D_{z} \bH$ is bounded.
The $t$-th component of $\bT$ is related to the $t$-th component
of $\bH$ by
\[
	H_t = T_t^2 - z_t^2.
\]
Differentiating this at the fixed point that exists according
to \Th{thrm:exist} gives 
\[
	D_{z} \bH = \diag(2z_t) ( D\bT - \id ).
\]
The operator $D_{z} \bH$ has bounded inverse because $z_t$ is bounded away 
from zero by \eq{BMdef} and the inverse of $D\bT-\id$ exists because 
$||D\bT|| < 1$.  \qed

This result can be extended to imply that the invariant set is 
uniformly hyperbolic for the case $b=1$ when the operator 
$D_{z}\bH$ is symmetric\cite{ABM92}.

In \Secref{sec:horse} we will use numerical 
continuation to estimate the parameters at which the first bifurcation 
occurs, giving an improvement in this bound, albeit a numerical one.

It is interesting that Devaney and Nitecki \cite{Devaney79} obtained 
precisely the same bound, \eq{epsmax}, for the parameter domain in which the 
non-wandering set of the \hen map is a hyperbolic horseshoe.  
Nevertheless the AI continuation argument has has two advantages 
over the geometrical arguments of Devaney and Nitecki.  First, it 
easily generalizes to higher dimensions allowing one to compute 
parameter bounds for the existence of horseshoes in higher dimensional 
maps \cite{Sterling98b}.  Second, it allows us to easily bound the 
parameter range for which certain subsets of orbits exist (i.e.  the 
parameter range where the map is conjugate to a subshift of finite 
type).  We present such a bound for a subshift of finite type in 
\Secref{sec:subshift}.


\Sec{Numerical Method}{sec:numerics}

In this section we formulate the problem of following \hen map orbits 
away from the anti-integrable limit as a classical continuation 
problem \cite{Keller}.

A period $n$ orbit family of the \hen map with coordinates given 
by $\bz(\epsilon)$
is a zero of the function $\bG : \mR^n\times \mR \rightarrow \mR^n$ 
whose $t^{\rm th}$ component is given by the left hand side of \eq{impmap}. 
The zeros of $G$ are generically smooth curves in $\mR^n\times \mR$ defined
by the continuation problem
\[ \label{contsystem}
     \left\{\begin{array}{lcl} 
        \bG(\bz,\epsilon) & = & {\bf 0} \\
        \bz(0)            & = & \bs 
     \end{array} \right. \;. 
\]
Practically the continuation is always started at the AI limit.
The curve might either extend to some maximal $\epsilon_{\rm max}$ 
and return the the limit or just continue indefinitely. Since
for the \hen map there are no orbits for $k<-1$ we expect that all
the curves do return to the AI limit.

This is a standard 
continuation problem, which we solve using a predictor-corrector 
method with a linear tangent predictor and a Newton's method corrector.  
For numerical linear algebra we use the Meschach library \cite{MESCHACH}.
The algorithm incorporates an adaptive step size control with bisection 
backtracking if the corrector fails to converge.  The algorithm 
terminates when a user-specified value of $\epsilon$ is reached or, 
when the tangent direction is not uniquely defined.  The process of 
continuing a sequence of orbits is trivially parallelizable since the 
operations performed on each orbit are completely independent of each 
other.  We use a ``divide and conquer'' strategy to spread the total 
computational effort across several different machines running 
simultaneously.  This is especially advantageous when the number of 
orbits continued reaches into the millions.

Continuation methods are based on the assumption that the orbits of 
interest are actually connected to the limit at which the continuation 
starts.  Since the \hen map does not have an integrable limit, the 
natural starting point would be to continue all the orbits that 
bifurcate off the fixed points.  But this would only yield a small 
fraction of all orbits: many of them are born in saddle-node 
bifurcations that are not connected to any other orbit.  Therefore the 
AI limit is a much better limit from which to continue.  However, the 
same general restriction applies, i.e.~only orbits that are connected 
to the limit (or their parents, grandparents etc.) can be found.  We 
formulate this central hypothesis as the ``no-bubble-conjecture'':
\begin{con} [No Bubbles]
For the area preserving \hen map every orbit is (at least) 
continuously connected to the anti-integrable limit.
\end{con}
We could only hope for continuous connection in $\mR^n\times \mR$ 
because the branches corresponding to the children in a bifurcation 
are in general not smoothly connected to the parent.  Our numerics 
currently does not perform any branch-switching from parents to 
children.  Therefore, in practice, we are actually using the working 
hypothesis:
        ``Every orbit of the \hen map can be {\em smoothly} 
        connected to the anti-integrable limit.''
Unfortunately our working hypothesis is not true in general.  For $b 
\not = -1,0,1$ it has been shown \cite{Yorke92} that periodic orbits 
are both created and destroyed when the map parameter $k$ is increased 
(the authors of \cite{Yorke92} call this ``antimonotonicity'').  
Moreover, in \cite{DMS99} we show that this conclusion holds for the 
cases $b=\pm 1$ as well.  Consequently the topological entropy is not 
necessarily a monotone increasing function, as it is for the logistic 
map, $b=0$.  In the following we will elucidate the relation between 
antimonotonicity, our working hypothesis and the no-bubble 
conjecture.

Even though we have antimonotonicity whenever $b\not = 0$, this does 
not readily imply that our working hypothesis is false.  In particular 
the smallest period orbit that is antimonotonic when $b=1$ is of 
period $10$, and it is still smoothly connected with the AI limit 
\cite{DMS99}. However, orbits that bifurcate from this one may violate 
the working hypothesis.\footnote{
	This orbit is created in a $3/10$ rotational bifurcation of the 
	elliptic fixed point, and it initially moves towards smaller $k$ 
	values.  While it is traveling in the ``wrong'' direction, the 
	elliptic $3/10$ orbit has a winding number that does not exceed 
	$1/5000$; therefore orbits that bifurcate from it are of period $50000$ 
	or higher.  We suspect that these would be orbits that are not 
	smoothly connected to the AI limit, and therefore violate our working 
	hypothesis.  Since they bifurcate off the $3/10$ orbit which in turn is 
	smoothly connected, they are, however, at least continuously connected 
	to the AI limit, so that the no-bubble conjecture could still be 
	true.
}

The worst possible case from the point of view of continuation is an 
orbit that neither smoothly nor continuously connects to the limit, i.e. 
an ``isolated bubble.''  
Note that in the area preserving case, one orbit of this type implies 
an infinite number of them because it must be born in a saddle node 
bifurcation and the stable orbit of the created pair generically 
passes through an infinite number of rational winding numbers.  In 
order to constitute a violation of our conjecture, none of these 
orbits would be allowed to reach the AI limit; otherwise the original 
orbit would be continuously connected.

In the dissipative case, the lowest period example we were able to 
find of a periodic orbit that is not smoothly connected occurs for 
$b=-0.46$, where at $k\approx 1.0346$ the period $8$ orbit with 
sequence $\per{-^5+-+}$ has a period doubling bifurcation that creates 
a period 16 orbit that is not smoothly connected to the AI limit.  
Since this is the start of a period doubling sequence resulting in a 
what appears to be a strange attractor at a slightly smaller value of 
$k$, we expect there are many saddle-node bifurcations in this region 
that create orbits as $k$ decreases that are (presumably) not 
connected to the AI limit. This is reason to believe that 
the ``no bubbles'' conjecture is false when $|b| < 1$.

In any case, continuation from the anti-integrable limit has the 
advantage that at the beginning point all orbits exist, and they all 
continue nondegenerately.


\Sec{Symbolic Dynamics}{sec:symb}

In this section we introduce some notation for symbol sequences and 
bifurcations.  For simplicity we concentrate mostly on the 
area preserving case, $b = 1$, though many results apply generally.

Orbits in the anti-integrable limit are bi-infinite sequences $\bs 
\in 
\Sigma$. When it is needed, we will indicate the current time along an 
orbit using a ``$\sdot$'', so that $\bs = \ldots 
s_{-2}s_{-1} \sdot s_{0}s_{1}s_{2}\ldots$.  The dynamics on $\bs \in 
\Sigma$ 
are given by the shift map, $\sigma: \Sigma \rightarrow \Sigma$ 
defined as
\[
\sigma (\ldots s_{-1} \sdot s_{0}s_{1}s_{2}\ldots) = 
        \ldots s_{-1}s_{0} \sdot s_{1}s_{2}\ldots
\]
An orbit of the symbolic dynamics is periodic if the 
sequence $\bs$ is periodic.  We will denote an orbit of least period 
$n$ by the string of $n$ symbols and a superscript $\infty$ to 
represent repetition:
\[
        \per{s_{0}s_{1}\ldots s_{n-1}} = 
        \ldots s_{n-2} s_{n-1} \sdot  s_{0} s_{1} \ldots s_{n-1}s_0\ldots
\]
Of course any cyclic permutation of a periodic orbit gives another 
point on the same orbit.

Trivially, the map $\sigma$ has two fixed points, $\per{+}$ and 
$\per{-}$, and these correspond to the two fixed points of the \hen map.  
These are born in a saddle-node bifurcation at $k = -(1+b)^{2}/4$, 
which we denote by
\[
  \bif{\sn}{ \per{+} ,\per{-}} \;.
\]
We denote bifurcations with the general template 
\[ 
   \ptcbif{parent}{type}{children} \;,
\]
where $parent$ refers to the orbit that is undergoing the bifurcation, 
if any, and $type$ is one of $\sn$, $\pf$, $\pd$, or $m/n$, 
corresponding to a saddle-node, pitchfork, period doubling, or 
rotational bifurcation, respectively.  The set of orbits created in 
the bifurcation is listed as the $children$. When the stability of 
these differ, we adopt the convention that the unstable child is 
listed first, and the stable one second.

When $b=1$ the fixed points of the \hen map are located at
\[
  z_{\pm} =  \pm \sqrt{1 +\epsilon^{2}}+\epsilon
          = \epsilon \left( 1  \pm \sqrt{1+k} \right) \; .
\]
The stability of a period $n$ orbit of an area preserving map $f$ is 
conveniently classified by the ``residue'' defined as
\[
   R  = \frac14 \left(2 - \Tr(Df^{n}) \right) \;,
\]
so that an orbit is hyperbolic if it has negative residue, elliptic 
when $0<R<1$ and is reflection hyperbolic when $R > 1$ \cite{Meiss92}.  
The residues of the fixed points are
\begin{equation} \label{resfixed}
   R_{\pm} = \mp \frac{1}{2\epsilon} 
\sqrt{1+\epsilon^{2}} = \mp \frac12 \sqrt{1+k} \;,
\end{equation}
so the sign of the symbol is opposite to the sign of the residue of 
the fixed point. Thus the orbit $\per{+}$ is always hyperbolic, while the 
orbit $\per{-}$ is reflection hyperbolic for small $\epsilon$, or 
large $k$, but becomes elliptic at $\epsilon = 1/\sqrt3$, or $k=3$.

\begin{table}[tb]
      \centering
      \begin{tabular}{c|c|cc|c}
 Parent                  & Type         & Child                 & Child                 & $k$-Value  \\  \hline 
                & $\sn$  & $\per{-}$    & $\per{+}$     & $-1$  \\ \hline
$\per{-}$                & $\pd$        &               & $\per{+-}$    & $3$   \\ \hline
                & $\sn$  & $\per{-+-}$          & $\per{+-+}$   & $ 1$  \\
$\per{-}$                & $1/3$        & $\per{-+-}$           &               & $ \frac54$ \\ \hline
$\per{-}$                & $1/4$        & $\per{+--+}$  & $\per{+-++}$  & $ 0$  \\ 
$\per{+-}$& $\pd$&              & $\per{-+--}$  & $ 4$  \\ \hline
$\per{-}$                & $1/5$        & $\per{-+++-}$         & $\per{++-++}$         & $\frac{7-5\sqrt5}{8}$ \\
$\per{-}$                & $2/5$        & $\per{--+--}$         & $\per{-+-+-}$         & $\frac{7+5\sqrt5}{8}$ \\
                & $\sn$  & $\per{+-+-+}$        & $\per{+---+}$         & $5.5517014^\dagger$ \\
  \hline
$\per{-}$        & $1/6$        & $\per{-+^{4}-}$       & $\per{++-+^{3}}$      & $-\frac34$    \\
$\per{+-+}$& $\pd$&             & $\per{+-^{4}+}$       & $\frac54$     \\
$\per{+-^4+}$   & $\pf$         & $\per{++-+--}$        & $\per{--+-++}$        & $3$   \\
                & $\sn$ & $\per{--+-+-}$        & $\per{--+-^3}$        & $3.7016569^\ddagger$ \\
$\per{+-}$      & $1/3$ & $\per{--+-+-}$        &               & $\frac{15}{4}$        \\
                & $\sn$ & $\per{+-+^{3}-}$      & $\per{--+^{3}-}$      & $5.6793695^\ddagger$ \\
  \hline
\multicolumn{5}{l}{$^\dagger  \, 16 k^5-108 k^4+105 k^3+27 k^2-97k-47$}\\
\multicolumn{5}{l}{$^\ddagger \, 16 k^6-136 k^5+213 k^4+220 k^3+126 k^2+108 k+81$}\\
    \end{tabular}
    \caption{Periodic orbits of the \hen map up to period 6 and their 
    bifurcations when $b=1$.  In the ``type'' column, ``$\sn$'' indicates 
    a saddle-node bifurcation, ``$\pf$'' a pitchfork bifurcation, and 
    ``$\pd$'' a period doubling bifurcation.  A rotational 
    bifurcation is denoted by $m/n$, referring to the winding number of 
    the parent at the bifurcation.  For $1/3$ the child is not created in 
    the bifurcation, it exists before and after the bifurcation.  If there 
    are two children, the one listed in the first column has negative 
    residue just after birth (except for the $\pf$ case).  The real roots 
    of the polynomials in the last rows give exact bifurcation values for 
    the three approximations shown.}
    \label{tbl:orbits}
\end{table}

For $b=1$ the sequence $\per{+-}$ corresponds to the period two orbit
\[
   \per{+-} \,:\, (z_{0},z_{1}) =  (\sqrt{1-3\epsilon^{2}}-\epsilon 
,-\sqrt{1-3\epsilon^{2}}-\epsilon) \;.
\]
This orbit exists only for $\epsilon < 1/\sqrt3$, and  is created by a
period doubling of the elliptic fixed point (when 
$R_{-}=1$).  We denote this bifurcation by
\[
    \ptcbif{\per{-}}{\pd}{ \per{+-} } \;.
\]
Similarly there are two period three orbits, 
\begin{eqnarray*}
        \per{--+} &:& (z_0, z_1, z_2) =  (-\sqrt{1-\epsilon^{2}}, 
            -\sqrt{1-\epsilon^{2}},\sqrt{1-\epsilon^{2}}-\epsilon) \\
        \per{-++} &:& (z_0, z_1, z_2) =  (-\sqrt{1-\epsilon^{2}}-\epsilon,
        \sqrt{1-\epsilon^{2}}, \sqrt{1-\epsilon^{2}}) \;.
\end{eqnarray*}
These are created in a saddle-node bifurcation at $k=\epsilon=1$;
\[  
    \bif{\sn}{ \per{--+} ,\per{-++} } \;.
\]

We list the low period orbits and their bifurcation values in 
\Tblref{tbl:orbits}.  Another class of bifurcations shown in the table 
are rotational bifurcations. A rotational bifurcation occurs when the
winding number of an elliptic orbit 
becomes $\omega = m/n$; we denote such bifurcations by the winding 
number of the parent orbit.  For example the birth 
of orbits with winding number $1/n$ at the fixed point 
$\per{-}$ is denoted
\begin{equation} \label{rotorbits}
        \ptcbif{\per{-}}{1/n}{\per{--+^{n-2}},\per{-++^{n-2}}} \;.
\end{equation}
This particular rotational bifurcation 
occurs when the multipliers of the fixed point are $e^{i2\pi\omega}$ 
or using \eq{resfixed}, when $k$ is given by
\begin{equation} \label{Romega}
          k_{\omega}= \cos({2\pi \omega})(\cos({2\pi \omega})-2) \;.  
\end{equation}
We have empirically identified the symbol sequences for rotational 
bifurcations, and will present the general symbolic formula for these 
and for rotational ``island around island'' orbits in 
\cite{Sterling98b}.

The residue of any periodic orbit of a Lagrangian 
system is easily 
computed from the matrix $M$ formed from the second variation of the 
action \cite{MacKay83}. For a period $n$ orbit of the \hen map this formula 
gives:
\[ 
     R(z(\epsilon))  =  -\frac14\frac{\det(M)}{\epsilon^{n}} \;,
\]
where $M$ is the periodic tridiagonal matrix with elements
\[  
  M_{t,t-1}= -b\epsilon \,, \quad
  M_{t,t} = 2z_t(\epsilon) \,, \quad
  M_{t,t+1} = -\epsilon \;.
\]
As we approach the anti-integrable limit, $\bz(\epsilon) \rightarrow \bs$
as $\epsilon \rightarrow 0$ and $M$ approaches the 
diagonal matrix $\mbox{Diag}(\-2 s_{i})$. Thus we see that the residue
becomes infinite at the anti-integrable limit and its sign is given by
$- \prod_{t=0}^{n-1}s_{t}$. Hence,
\begin{equation}
\label{residue}
       \sign(R(\bs)) = -(-1)^{j} \;,
\end{equation}
where $j$ is the number of minus signs in the symbol sequence $\bs$.


\Sec{A Subshift of Finite Type}{sec:subshift}

In this section, we extend \Th{thrm:exist} to the case of a subshift 
of finite type.  In particular, the biggest restriction in the proof of the 
theorem arises from the fact that the lower bound on the operator $\bT$ 
given in \eq{Tdef} is weakest when the signs $s_{i+1}=s_{i-1}=-1$ 
for positive $b$.  
We can improve the bound by restricting the set of admissible symbol 
sequences to forbid this particular case.  The shift map restricted to 
this subspace is a subshift of finite type with the forbidden set 
$\cal{F}=\{-+-,-\,-\,-\}$; that is, we define the shift space
\[ 
      \sigmaF = \Sigma \setminus \{ \bs : 
      \exists \; t \in \mZ \mbox{ such that } s_{t-1}=s_{t+1}=-\}  \; .
\]
This subshift can be easily described as a subshift on $2$-blocks 
represented by $\{--,-+,+-,++\}^{\mZ}$.
The subshift on the two-block space is represented by a 
vertex graph with the state transition matrix \cite{LM95}
\[
    S = \left(\begin{array}{cccc} 
                0 & 0 & 1 & 0 \\
                1 & 0 & 1 & 0 \\
                0 & 0 & 0 & 1 \\
                0 & 1 & 0 & 1
         \end{array} \right) \;,
\]
which indicates by $S_{ij} = 1$ an allowed transitions from state $j$ to 
state $i$. 
The two zeros $S_{11}$ and $S_{32}$ come from the forbidden
sequence in $\sigmaF$, the remaining are obtained because
successive two-blocks overlap in one symbol, i.e.,\ 
$(s_{t-1}s_{t})$ has to be followed by $(s_{t}s_{t+1})$.  
The number of fixed points of period $n$ for the subshift is given by
\[
  \Tr(S^{n}) = \gamma^{n}+(1-\gamma)^{n} +2 (-1)^{n/2}(n-1 \mod 2) \;,
\]
where $\gamma = (1+\sqrt{5})/2$ is the golden mean. Thus the 
topological entropy for $\sigmaF$ is $ \ln \gamma$.
The number of distinct periodic orbits can be obtained 
from the trace formula by subtracting the number of periodic orbits 
for all factors of $n$ and then dividing by the number
of cyclic permutations, $n$.  For 
comparison with the full shift and with the numerical results below, 
we give a list of these in \Tblref{tbl:periods}.  For example 
there are a total of $1,465,020$ periodic points of the full shift 
with period $n \le 24$, while there are only $12,216$ in the subshift 
$\sigmaF$.
{\small
\begin{table}[tbp]
	    \centering
	\begin{tabular}{r|r|r}     
		Period   & \multicolumn{1}{c}{$\Sigma$}&\multicolumn{1}{c}{$\sigmaF$} \\
		\hline
		 1 & 2         & 1 \\
		 2 & 1         & 0 \\
		 3 & 2         & 1 \\
		 4 & 3         & 2 \\
		 5 & 6         & 2 \\
		 6 & 9         & 2 \\
		 7 & 18        & 4 \\
		 8 & 30        & 5 \\
		 9 & 56        & 8 \\
		10 & 99        & 11\\
		11 & 186       & 18\\
		12 & 335       & 25\\
	\end{tabular}
	\hspace*{5pt}
	\begin{tabular}{r|r|r}     
		Period   & \multicolumn{1}{c}{$\Sigma$}&\multicolumn{1}{c}{$\sigmaF$} \\
		\hline
		13 & 630       & 40\\
		14 & 1161      & 58\\
		15 & 2182      & 90\\
		16 & 4080      & 135  \\
		17 & 7710      & 210  \\
		18 & 14532     & 316  \\
		19 & 27594     & 492  \\
		20& 52377     & 750  \\
		21& 99858     & 1164 \\
		22& 190557    & 1791 \\
		23& 364722    & 2786 \\
		24& 698870    & 4305 \\
	\end{tabular}
	        \caption{Number of orbits with minimal period $n$ of the 2-shift 
		and the subshift $\sigmaF$.} 
	\label{tbl:periods}
\end{table}
}

When $b$ is non-negative, orbits with symbol sequences in the subspace 
$\sigmaF$ can be shown to persist longer than a general orbit:

\begin{teo}[Existence and Uniqueness of $\sigmaF$ orbits] 
\label{thrm:seq}
        Suppose $0 \le b \le 1$.  For every symbol sequence $\bs \in 
        \sigmaF$ there exists a unique orbit $\bz(\epsilon)$ of the 
        \hen map \eq{henmap} such that $\bz(0) \equiv \bs$ providing $0 \le 
        \epsilon < \epsilon_{\rm max}$, where
        \begin{equation}
             \epsilon_{\rm max} \equiv \frac{2}{1+b}\sqrt{\frac{-b^2+2b+5
              - 2\sqrt{5+4b}}{(1-b)(5-b)}} \;.
        \end{equation}
\end{teo}
\noindent
This theorem follows from the same argument that gave 
\Th{thrm:exist} with only minor modifications.  We summarize the changes 
in the argument in the following discussion.

{\bf Proof:} When $0 \le b \le 1$, $\bs \in \sigmaF$, and $\bz \in B_{M}$, 
we can bound the norm of iterates of $\bT$ in \eq{Tdef} using the 
inequalities
\[
  \alpha_{k} \le ||\bT^{k}(z) ||_{\infty} \le \beta_{k} \;,
\]
where the coefficients $\alpha_{k}$ and $\beta_{k}$ are determined by 
the recursions
\begin{eqnarray*}  
\beta_{k} &=& \sqrt{1 +\epsilon(1+b)\beta_{k-1}} \;, \\
\alpha_{k}&=& \sqrt{1 +\epsilon (b\alpha_{n-1} - \beta_{n-1})}  \;,
\end{eqnarray*}
with $\beta_{0}= 1+M$ and $\alpha_{0}= 1-M$. The sequence $\beta_k$ is 
the same as that in \cite{Sterling98a}; it has the unique attracting 
fixed point
\[ 
    \beta_\infty =\frac12 \left\{\epsilon(1+b) + 
\sqrt{\epsilon^2(1+b)^2 + 
    4}\right\}  \;. 
\]
Since the recursion for $\{\alpha_n\}$ depends on $\beta$, but not 
vice versa, the coupled system also has a unique attracting fixed 
point, which is given by $(\alpha_{\infty},\beta_{\infty})$ with
\[ 
      \alpha_\infty = \frac12 \left\{\epsilon b + \sqrt{\epsilon^2b^2
     +4(1-\epsilon\beta_\infty)}\right\}  \;.
\]
This implies that for large enough $n$, $\bT^{n}$ maps the ball 
$B_{1-\alpha_{\infty}}$ into itself.  The map $\bT$ is a contraction 
map on this ball providing $||D\bT^k||_{\infty} < 1 $.  This leads to 
the same bound as that in \cite{Sterling98a}, namely:
\[
   \epsilon_{\rm max}(1+b) < 2\alpha_\infty\;.
\]
After some simplification, this inequality yields the formula for 
$\epsilon_{\rm max}$. 
The condition that the operator $\bT$ be real, 
$\epsilon < 1/(\beta_\infty - b \alpha_\infty)$, is satisfied whenever
the map is a contraction.
\qed

Similar arguments for negative $b$ lead to bounds for a 
subshift with the forbidden subsequence $+\ast -$
(where $\ast$ is any symbol). However, this subshift is not of
much interest, since there are only three periodic orbits in it.


\Sec{Horseshoe Boundary}{sec:horse}

Theorem \ref{thrm:exist} provides an analytical bound on 
the parameter range for which the \hen map has a hyperbolic horseshoe.  
This bound corresponds to the dark grey region in 
\Fig{fig:horsebound}.  
According to \Th{thrm:seq} the subshift $\sigmaF$ exists, in addition,
in the lighter shaded
area in the figure.
This bound is valid only for $b \ge 0$, and meets the former at $b=0$.
 
Here, we use our continuation method to estimate the boundary of 
existence of the horseshoe by following all orbits up to 
period 24 from the  anti-integrable limit.
In order for the numerical boundary to be valid we only need to
assume that we can extrapolate from 24 to $\infty$. Since 
there are at most $2^n$ periodic 
points of period $n$ in the \hen map \cite{Moser60}, we know that we are not
missing any orbits because we have them all at the AI limit.
This is no longer true after the first bifurcation
because orbits that have disappeared might reappear
for smaller $k$.

\Epsfig{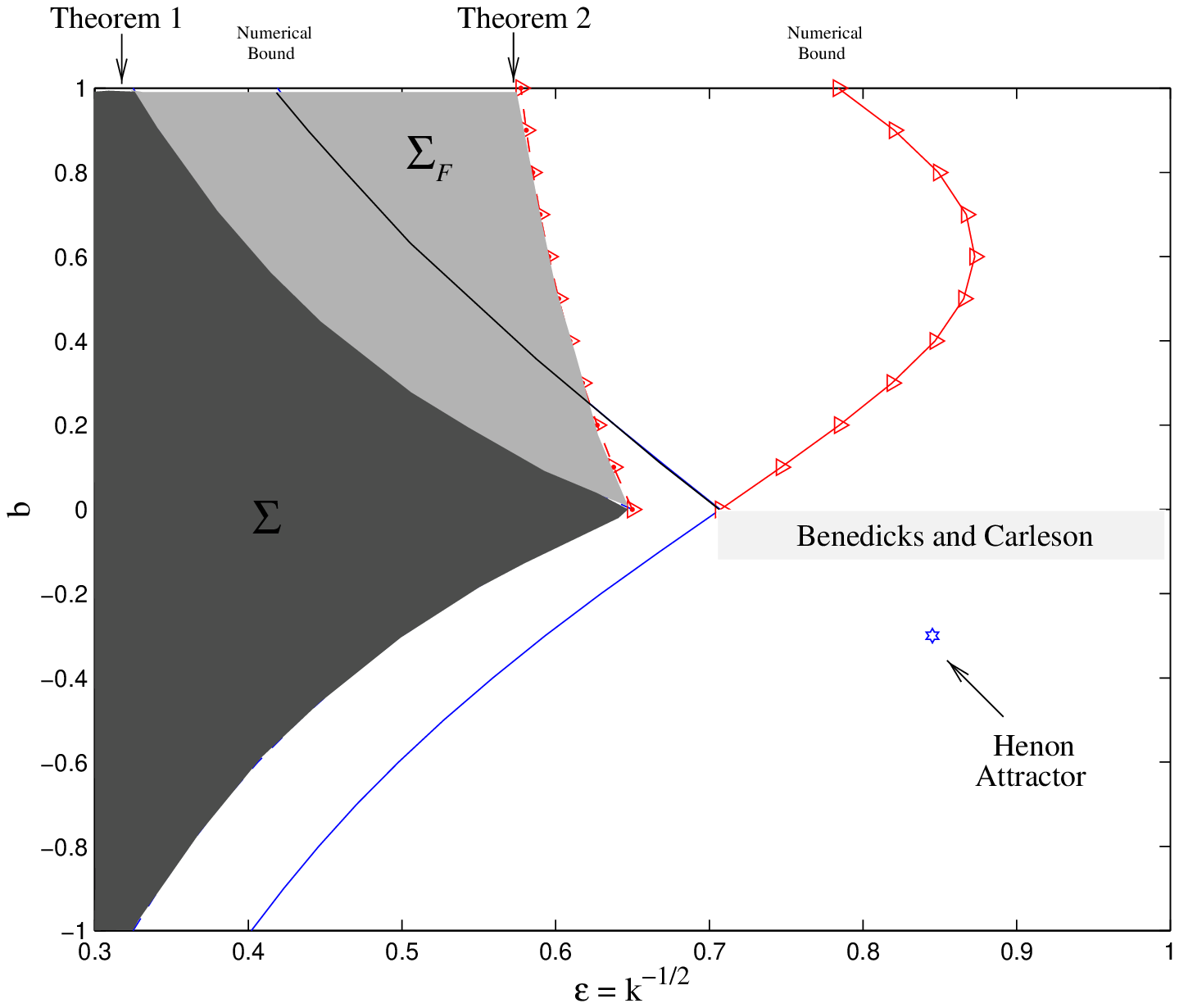}{tbp}
        {First bifurcations for the \hen Map.  The dark shaded region 
        represents \Th{thrm:exist} and the lighter that of \Th{thrm:seq}.  
        The curves represent the numerical results for the first orbits 
        destroyed up to period $24$.  Bounds for the subshift 
        $\sigmaF$ are indicated with a triangle symbol.} 
        {fig:horsebound}{3.5in}

To construct a numerical approximation for the boundaries, we first 
generate all symbol sequences for orbits of periods up to $24$.  
Then, for fixed $b$, we numerically continue each orbit in $\epsilon$ 
away from the anti-integrable limit and monitor its multipliers to 
detect bifurcations.  For each $b$ we record the smallest value of $\epsilon$ 
at which a bifurcation occurs.  The resulting numerical bounds 
in \Fig{fig:horsebound} are shown as solid curves.

The numerical bound for the full shift is similar in shape to the 
analytical one, but shifted to the right in $\epsilon$.  While the 
analytical bound is symmetric under $b \rightarrow -b$, the numerical 
results are not.  For example the first bifurcation at $b=1$ occurs 
for $\epsilon \approx 0.41888$, while at $b=-1$ it occurs for 
$\epsilon \approx 0.40167$.  In the logistic limit ($b=0$), 
\eq{henmap} reduces to the logistic map,
\begin{equation}\label{logistic}
    z_{t+1}= \frac{1}{\epsilon}(z_{t}^{2}-1) \;,
\end{equation}
for which the first bifurcation occurs at $\epsilon = 1/\sqrt2$, 
where the orbit of the critical point becomes bounded.

When $b$ is positive, the symbol sequences for the first pair of 
orbits destroyed up to period $24$ extrapolate to orbits that are 
homoclinic to the fixed point $\per{+}$; we conjecture that these are 
the first orbits destroyed as $\epsilon$ increases from $0$:

\begin{con} \label{con:firstbif}
   For positive $b$, the first bifurcation as $\epsilon$ 
   increases from $0$ corresponds to the homoclinic saddle-node bifurcation
   \begin{equation} \label{firstbif} 
      \bif{\sn}{+^\infty-(+)-+^\infty, +^\infty-(-)-+^\infty} \;. 
   \end{equation}
\end{con}

The parenthesis in the middle enclose the ``core'' of the homoclinic 
orbit, see the next section.  A theorem of Smillie \cite{Smillie97} 
implies that the first bifurcation destroying the \hen horseshoe must 
be a quadratic homoclinic tangency for {\it some} orbit.  Our 
observations imply that it is a homoclinic bifurcation of $\per{+}$.  
When $b<0$, however, the most natural description of the first 
bifurcation is as a heteroclinic tangency, which leads to

\begin{con}
   For negative $b$, the first bifurcation as $\epsilon$ 
   increases from $0$ corresponds to the heteroclinic saddle-node bifurcation
   \[ 
      \bif{\sn}{ -^\infty+(-)-+^\infty ,  -^\infty+(+)-+^\infty } \;. 
   \]
\end{con}

This does not contradict Smillie's theorem, as there are many homoclinic
bifurcations that accumulate on this heteroclinic bifurcation. For example,
for each $m$ the orbits $(-^{m}++-+^m)^{\infty}-+(-^{m}++-+^m)^{\infty}$,
are homoclinic to the periodic orbit $(-^{m}++-+^m)^\infty$, and the
bifurcation points of these homoclinic orbits limit on that of 
the heteroclinic orbits as $m \rightarrow \infty$.

Filtering the symbol sequences to choose only those in the subshift 
$\sigmaF$, we can use the same numerical data previously described
to find the first 
bifurcation amongst the orbits in $\sigmaF$. This gives the solid 
curve marked with triangles in \Fig{fig:horsebound}. This curve 
has a qualitatively different shape than the analytical bound.

For reference we indicate in \Fig{fig:horsebound} the point $k=1.4$, 
and $b=-0.3$, corresponding to the much studied \hen  attractor. 
We also sketch the parameter range ($b$ small enough, $1<k<2$) for which 
the theorem of Benedicks and Carleson \cite{Benedicks91} implies that 
the \hen map has a transitive attractor with positive Lyapunov 
exponent.

Note that the numerical horseshoe boundary does {\em not} depend
on the no bubble conjecture. This is so because it is known that
there can be at most $2^n$ periodic points of period $n$ \cite{Moser60}.
Since we follow all of them up to period 24 there can be no other 
orbits up to that period. In other words, orbits first have to 
be destroyed before they can be reborn.


\Sec{Homoclinic Orbits}{sec:homo}

In this section we use the symbolic dynamics to classify orbits 
of the \hen map that are homoclinic to the hyperbolic fixed point $p = 
\per{+}$ and study their bifurcations.  We begin 
with some general terminology, referring to the \hen map as an example.

Let $f$ be an orientation preserving map\footnote
   {
    The orientation reversing case could be included by 
    considering $f^{2}$, since its manifolds have the same geometry 
    as those of $f$.
    }
of the plane with hyperbolic fixed point $p$.  The stable and 
unstable manifolds of $p$ are denoted by $\Wu$ and $\Ws$, and a 
closed segment of such a manifold between two points $\alpha$ and 
$\beta$ by $\Wu[\alpha,\beta]$.  We use a parenthesis to denote an 
open endpoint of a segment.  A segment that extends to the fixed 
point, e.g.  $\Wu(p,\alpha]$, is called an {\it initial segment} of 
the manifold.  The set of homoclinic orbits is the set of 
intersections $\Ws \cap \Wu$.  A point $\alpha$ is on a {\it primary} 
(or principal) homoclinic orbit if the two initial segments to 
$\alpha$ touch only at $\alpha$, i.e.,
\[
    \Wu(p,\alpha] \cap \Ws(p,\alpha] = \{\alpha\} \;.
\]
Thus the initial segments to a primary homoclinic orbit define a 
Jordan curve; we call the interior of this curve a resonance zone.  
More generally, a {\it resonance zone} is a region bounded by 
alternating initial segments of stable and unstable manifolds 
\cite{MacKay87,Easton98}.

For example, in \Fig{fig:tan} we sketch the left-going branches of the 
manifolds from $p = \per{+}$ for the area preserving 
\hen map.  There are precisely two primary homoclinic orbits; in the 
figure, we label points on these orbits with $\alpha$ and $\zeta$.  We 
choose to use $\alpha$ to construct the resonance zone.

\Epsfig{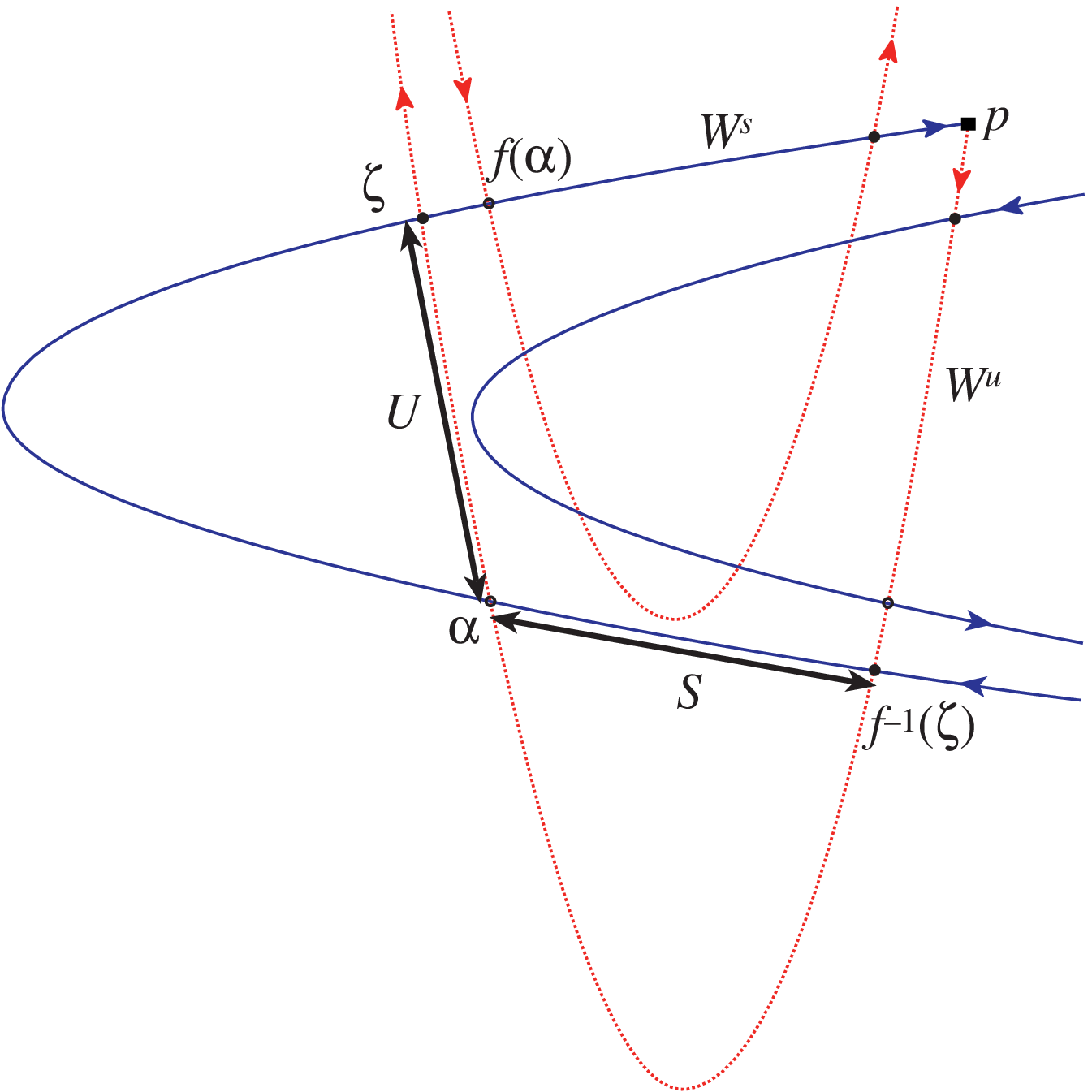}{tbp} {Stable and unstable manifolds for the \hen 
        map at $k=5$ and $b=1$ shown in $(z,z')$ coordinates.} 
        {fig:tan}{3in}

The stable manifold is divided into two invariant branches by the 
fixed point.  An ordering is defined on each branch of $\Ws$, so that 
$\beta <_{s} \gamma$ for two points on a branch of $\Ws$ if $\beta$ 
is nearer to $p$ along $\Ws$ than $\gamma$, i.e., $\beta \in 
\Ws(p,\gamma)$.  We similarly define an ordering $<_{u}$ on each 
branch of $\Wu$.

A segment of a manifold from a point to its iterate, 
$\Ws(\beta,f(\beta)]$, is called a {\it fundamental segment} 
\cite{Easton98}.  The union of the iterates of a fundamental segment 
is the entire branch of the manifold that contains $\beta$.  
Moreover, since the iterates 
are disjoint, every homoclinic orbit on this branch must have
precisely one point on each fundamental segment.

For the \hen map, we focus on the left-going branches of $\Ws$ and 
$\Wu$ and the fundamental segments between $f^{-1}(\zeta)$ and 
$\zeta$.  These also form the boundaries of the incoming and exit sets 
for the resonance zone defined by $\alpha$ \cite{Meiss97}.  The 
exterior halves of these segments, $\Ws(\alpha,\zeta)$ and 
$\Wu(f^{-1}(\zeta),\alpha)$, contain no homoclinic points since orbits 
on these segments are unbounded, so it is sufficient to look for 
homoclinic points on the interior halves,
\begin{eqnarray} 
    U &\equiv& \Wu[\alpha,\zeta] \;, \nonumber \\ 
    S &\equiv& \Ws[f^{-1}(\zeta),\alpha] \;.\label{SandU}
\end{eqnarray}
Every homoclinic orbit must have exactly one point on both $S$ and $U$.

Homoclinic orbits can be classified in a number of ways.  The {\it 
type} \cite{Easton98}, of a homoclinic point $\beta$ is\footnote
{
      Our definition of the type differs from Easton's slightly, to 
      comply with his definition that type 1 is equivalent to the 
      horseshoe. Rom-Kedar \cite{Rom92} uses the term {\it Birkhoff 
      signature} instead of type.
}
\[
   \type(\beta) = \sup \{j \ge 0 : \Ws(p,f^{j}(\beta)] \cap 
                       \Wu(p,\beta] \ne \emptyset \} \;;
\]
i.e., the number of iterates for which the stable initial segment to 
$f^j(\beta)$ intersects with the unstable initial segment to $\beta$.  
The type of a homoclinic point is invariant along its orbit.  Primary 
homoclinic points have type 0.

Homoclinic orbits on particular branches of $\Ws$ and $\Wu$ can also 
be classified by their {\it transition time}.  In general this is 
defined relative to a choice of a primary homoclinic point, $\zeta$ 
and the fundamental segments $\Wu(f^{-1}(\zeta),\zeta]$ and 
$\Ws(f^{-1}(\zeta),\zeta]$.  Any homoclinic orbit on these branches 
has exactly one point, $\beta$, on the unstable segment.  The 
transition time is the number of iterates required for 
$\beta \in \Wu(f^{-1}(\zeta),\zeta]$ to reach the stable segment:
\[
      t_\trans(\beta) = k \mbox{ if } 
                         f^{k}(\beta) \in \Ws(f^{-1}(\zeta),\zeta]
\]
The value of the transition time depends upon the choice of 
fundamental segments, so it is not as basic a property as the type.

In the simplest case, the transition time is easily related to the 
type of the orbit \cite{Easton86}:
\begin{lem} \label{typelem}
   Assume there are exactly two primary homoclinic orbits, $\zeta$ and 
   $\alpha$, and the segments $S$ and $U$ defined in \eq{SandU} 
   contain all of the homoclinic orbits.  Then for each homoclinic 
   point in $\beta \in U$, $t_\trans(\beta) = \type(\beta)$.
\end{lem}   
{\bf Proof:} If $\beta \in U$ is of type $t$, 
then by definition $\Ws(p,f^{t}(\beta)] \cap \Wu(p,\beta] \ne 
\emptyset$.  Now since $\alpha <_{u} \beta <_{u} \zeta$ and 
$W^{s}(p,\zeta) \cap W^{u}(p,\zeta) = \emptyset$, this implies 
that $\alpha <_{s} f^{t}(\beta)$.  However, $\Ws(p,f^{t+1}(\beta)] 
\cap \Wu(p,\beta] = \emptyset$, which means that $f^{t+1}(\beta) 
<_{s} \alpha$, but there are no homoclinic points on 
$\Ws(\zeta,\alpha)$, so actually $f^{t+1}(\beta) <_{s} \zeta$.  Now 
$S$ contains every homoclinic point that reaches $W^{s}(p,\zeta)$ in 
one iteration, so $f^{t}(\beta) \in S$. \qed

Each homoclinic orbit has a Poincar\'e signature that determines the 
direction of crossing of $\Wu$ and $\Ws$ at points on the orbit.  We 
define the signature to be $+1$ if, looking along the unstable 
manifold in the direction of motion, the stable manifold crosses the 
unstable from the left to the right side.  Crossings in the opposite 
direction have signature $-1$.  If the manifolds do not cross but
only touch (a topologically even intersection), the signature is defined 
to be zero.  Since the map is orientation preserving, the signature is 
invariant along an orbit.  Thus in \Fig{fig:tan} $\alpha$ and $\zeta$ 
have signatures $-1$ and $+1$, respectively.  The signature of a 
particular homoclinic orbit is typically not preserved in a 
bifurcation, but the total signature of the bifurcating orbits must be 
the same on each side of the bifurcation value.  For example a 
saddle-node bifurcation creates a zero signature orbit that splits 
into one positive and one negative signature orbit.

For the \hen map, the AI symbol sequence can be used for the
classification of homoclinic orbits.  It is easy to construct 
homoclinic and heteroclinic orbits using the symbolic dynamics: an 
orbit heteroclinic from a periodic orbit $\per{s}$ to a periodic orbit 
$\per{s'}$ has a symbol sequence that begins with a head sequence 
$\per{s}$ and ends with a tail sequence $\per{s'}$ with some 
arbitrary, finite symbol sequence separating the head and tail.  For 
example, the simplest orbits homoclinic to $p = \per{+}$ are the 
primary homoclinic orbits:
\begin{eqnarray}
    \zeta   &=& +^{\infty} - \sdot +^{\infty} \;, \nonumber \\
    \alpha  &=& +^{\infty} - \sdot - +^{\infty} \;,
\end{eqnarray}
corresponding to those we labeled in \Fig{fig:tan}.  These symbol 
sequences arise because as $\epsilon \rightarrow 0$ the point $\alpha$ 
moves to the point $ -\sdot-$, while $\zeta$ moves to $-\sdot+$ and 
$f^{-1}(\zeta)$ to $+\sdot-$.

All other orbits homoclinic to $p$ can be written in the form 
$+^{\infty}-(s) - +^{\infty}$, where $s$, the {\it core}, is any 
finite sequence---thus there is a one-to-one correspondence between 
finite symbol sequences and potential homoclinic orbits (all of which 
exist in the AI limit).  This implies, for example, that near the 
anti-integrable limit there are $2^{k}$ homoclinic orbits with core 
length $k$.  We will often denote an orbit homoclinic to $p$ 
simply by writing the core in parenthesis, $(s)$. Note that a given
core $(s)$ is not equivalent to any core with the same $s$ cyclically 
permuted.

The classification of homoclinic orbits by their symbol sequence can 
be used to compute other invariants.  To determine the type of an 
orbit, we simply note that the AI symbols give exactly the same coding 
for an orbit as the standard symbolic coding for the horseshoe.  This 
implies that the point $+^{\infty}-\sdot (s)-+^{\infty}$ corresponds 
to a phase point on $U$, and the point $+^{\infty}-(s)\sdot 
-+^{\infty}$ is on $S$, thus
\begin{lem} \label{transitAIL}
     The transition time of the homoclinic orbit close to the AI limit
     is given by the length of the core sequence. 
\end{lem}
\noindent
For example, the homoclinic orbit $+^{\infty}-(-\,-+)-+^{\infty}$ has 
the core sequence $(-\,-+)$, and therefore has transition time $3$.

Similarly the signature of a homoclinic orbit in the horseshoe
is given by simply counting the number of $-$ signs in the core sequence.
\begin{lem}
     The signature of a homoclinic orbit with core $(s)$
     close to the AI limit is given by $-(-1)^{j}$ where $j$ is the 
     number of $-$ signs in $s$.
\end{lem}
\noindent
Thus the orbit $(-\,-+)$ has signature $-1$.\footnote
   {\eq{residue} implies that the signature is the same as the limiting 
   sign of the residue of periodic orbits that approximate the homoclinic 
   orbit.}
We will see that, when $b=1$, some homoclinic orbits undergo pitchfork 
bifurcations, which change their signature, so this rule is not valid 
for all parameter values.

\Epsfig
        {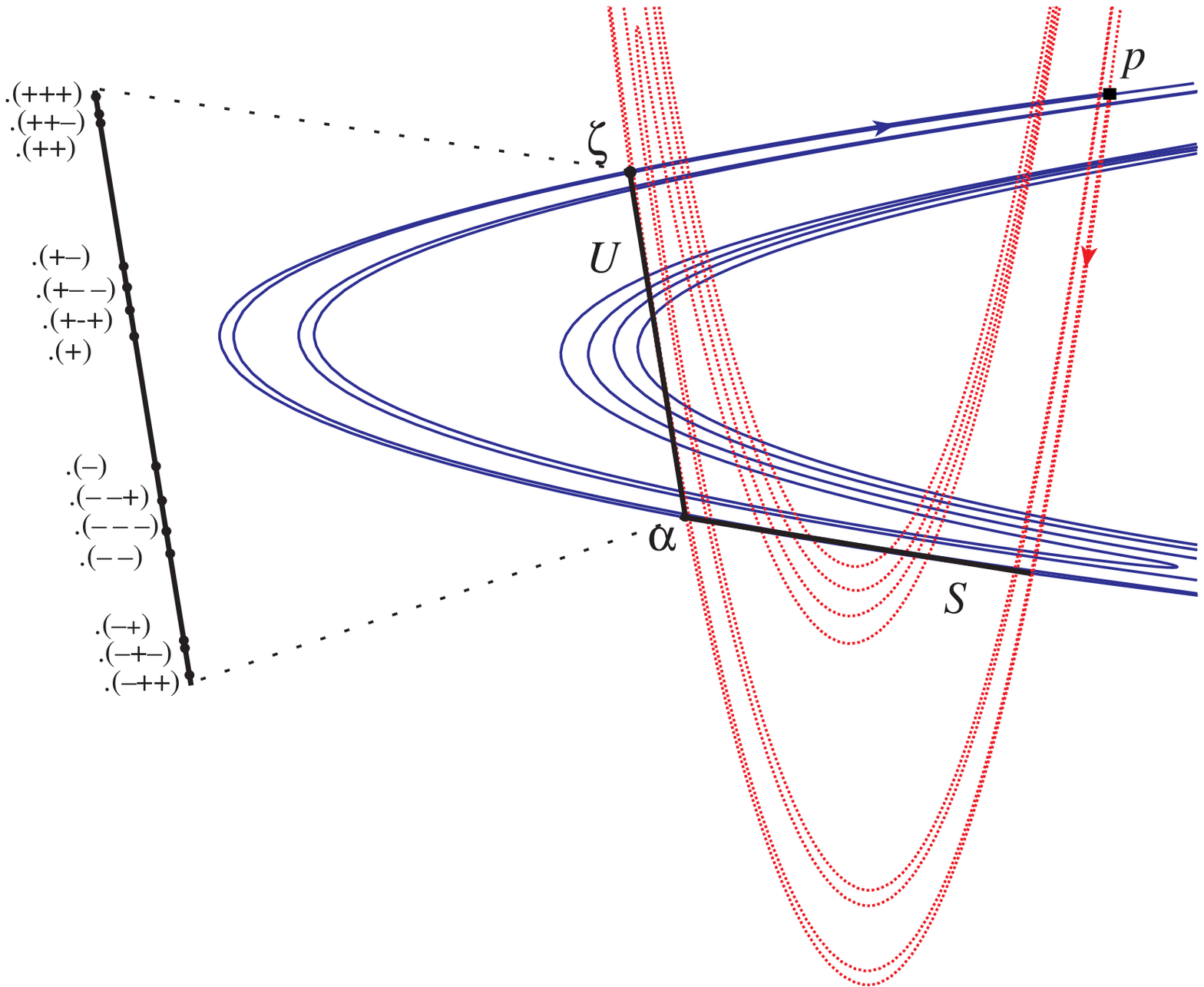}{ptb}
        {Ordering of homoclinic orbits of type 1, 2 and 3 at $k=6.25$.  The 
        enlargement on the left shows the core sequences for the $14$ 
        homoclinic orbits up to type $3$ The type 0 orbits $\alpha$ and 
        $\zeta$ are not listed.}
        {fig:homorder}{4in}

The positions of the homoclinic orbits on $U$ for orbits of type 1, 2 
and 3, labeled by their core sequences, are shown in 
\Fig{fig:homorder}.  
The order of the homoclinic orbits on the 
segments $S$ and $U$ close to the AI limit
must be the same as the corresponding ordering of homoclinic points 
in the complete horseshoe by continuity.
This ordering is equivalent to that of the logistic map, 
\eq{logistic}, for the orbits forward asymptotic to the $+$ fixed 
point.  This gives an easy way to compute the ordering, see 
\Fig{fig:logistic}. In the logistic limit, all of the sequences 
forward asymptotic to the fixed point are destroyed when the orbit of 
the critical point becomes bounded at $\epsilon = 1/\sqrt2$.  The 
fixed point $\sdot +^{\infty}$ has a single preimage, which is $\alpha 
= \sdot -+^{\infty}$.  Every other orbit that is forward asymptotic to 
$\sdot +^{\infty}$ has the form $\sdot (s)-+^{\infty}$.

\Epsfig
        {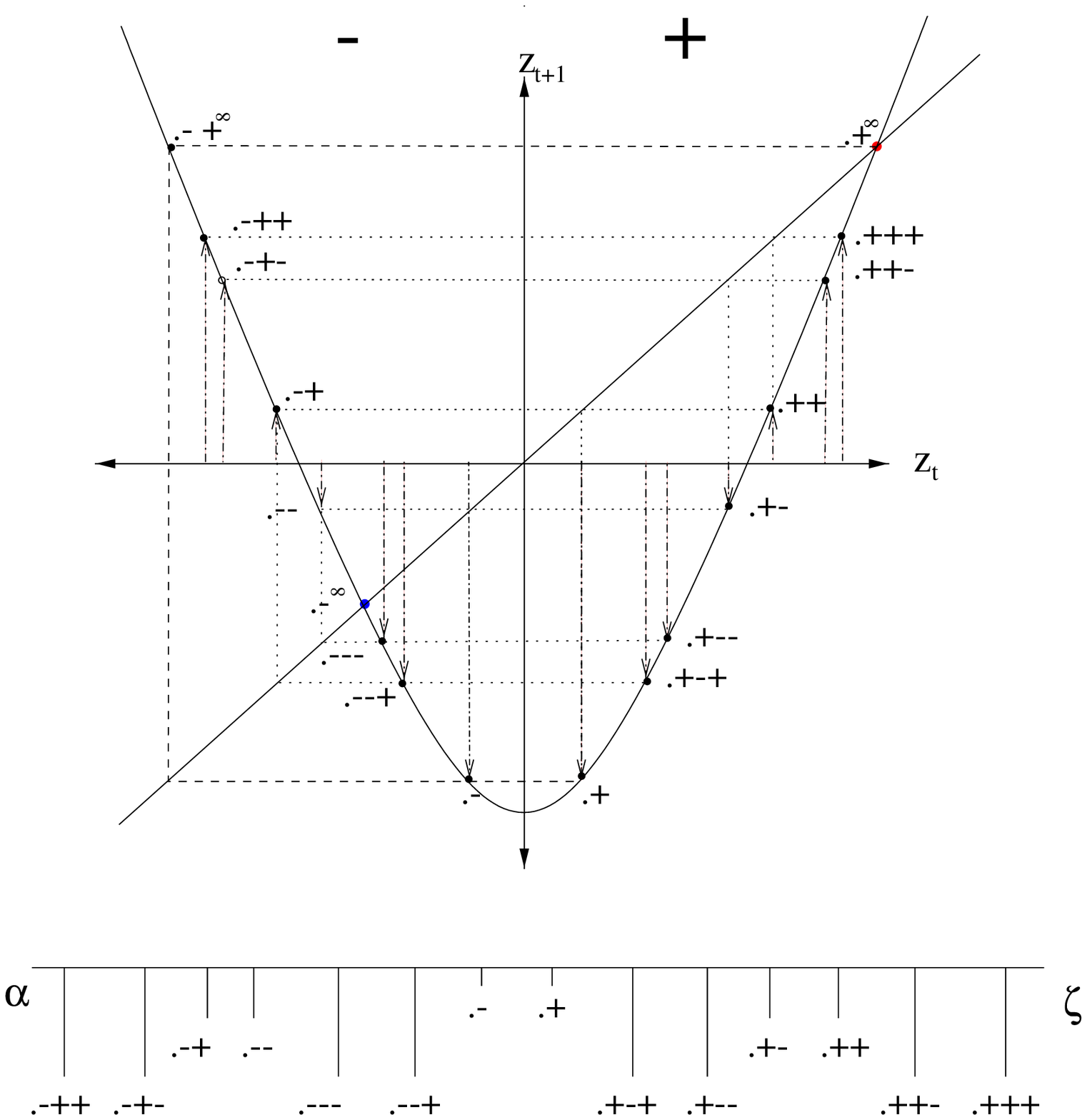}{tb} {Ordering of pre-periodic points for the $+$ fixed 
        point of the logistic limit of the \hen map. The symbols are 
        determined by the 
        itinerary of the orbit relative to the critical point at $z=0$.} 
                {fig:logistic}{4in}

In the area preserving case the ordering of the orbits along $S$ 
is equivalent to that on $U$ upon time reversal.  
Thus a type $t$ point $+^{\infty}-(s_{1}s_{2}\ldots 
s_{t}) \sdot -+^{\infty}$ on $S$ is in the same relative position as 
the point $+^{\infty}-\sdot( s_{t}s_{t-1}\ldots s_{1})-+^{\infty}$ on 
$U$. Close to the AI limit we always have this ordering on the manifolds,
which is just another way of saying that the map is conjugate to 
the horseshoe map.

So long as there are no homoclinic bifurcations, then the orderings 
$>_u$ and $>_s$ are just given by the usual unimodal ordering as stated in
\begin{lem} \label{orderAIL}
    The ordering $>_u$ on $U$ and $>_s$ on $S$ close to the AI limit
    is given by
\begin{eqnarray*}
        +^\infty-\sdot e+\dots & >_u & +^\infty-\sdot e-\dots \\
        +^\infty-\sdot o+\dots & <_u & +^\infty-\sdot o-\dots \\
        \dots+e \sdot-+^\infty & >_s & \dots-e \sdot-+^\infty \\
        \dots+o \sdot-+^\infty & <_s & \dots-o \sdot-+^\infty \; ,
\end{eqnarray*}
where $e$ / $o$ are finite sequences with an even / odd number of 
minus signs, respectively.
\end{lem}
The ordering shown in \Fig{fig:homorder} and \Fig{fig:logistic} 
is exactly this one upon appending the ``homoclinic tail'' $-+^\infty$
to the cores. The maximal orbit on $U$ is $\zeta$, corresponding
to the tail $\sdot +^\infty$, the minimal orbit is $\alpha$, corresponding to 
$\sdot -+^\infty$.

\Sec{Homoclinic Bifurcations}{sec:hobs}

Homoclinic bifurcations are bifurcations between homoclinic orbits. 
Compared to ordinary bifurcations of periodic orbits they possess
additional structure because the invariant manifolds 
(with their ordering)  must be involved in the bifurcation process.
To make this explicit, we say that
two homoclinic orbits $\beta$ and $\gamma$ are 
{\it double neighbors} if the segments $\Wu[\beta,\gamma]$
 and $\Ws[\beta,\gamma]$ contain no other homoclinic orbits.
Three ordered homoclinic points $\beta <_u \gamma <_u \delta$ are
{\it triple neighbors} if both $\beta, \gamma$ and $\gamma, \delta$ are
double neighbors.
An obvious observation with nevertheless important consequences
is the ``double neighbor'' lemma:
\begin{lem} \label{lem:doubleN}
        Two homoclinic orbits $\beta$ and $\gamma$ cannot bifurcate
        unless they are double neighbors.
\end{lem}
The converse gives a simple forcing relation: before $\beta$ and $\gamma$
can bifurcate any homoclinic orbit on either segment between them must
have disappeared.

Another consequence is the transition time lemma:
\begin{lem} \label{lem:ttime}
        If two homoclinic orbits $\beta$ and $\gamma$ bifurcate then they
        must have the same transition time $t_\trans$.
\end{lem}
{\bf Proof:}
Let $\beta$ and $\gamma$ be neighbors on $U$. If their transition time is
different then they are not neighbors on $S$, so they cannot bifurcate. \qed

This allows us to extend Lemma~\ref{transitAIL} away from the AI limit, so that
one can take the transition time as an adequate replacement of the period:
\begin{cor}
        The transition time of a homoclinic orbit never changes.
\end{cor}
{\bf Proof:}
Since the transition time is an integer it cannot change under
smooth deformations. It could only change at bifurcations,
but we have just seen that only orbits with the same transition time 
bifurcate. \qedd \\

Therefore homoclinic bifurcations only take place between double neighbors
with the same transition time, i.e., core length.
Close to the AI limit the horseshoe is still complete. In this situation it
is possible to find all neighbors:
\begin{lem} \label{lem:neighbor}
        Two homoclinic orbits on $U$ are neighbors in the complete horseshoe
        if and only if they are of the form 
        $+^\infty-\sdot(s+)-+^\infty$ and
        $+^\infty-\sdot(s-)-+^\infty$.
\end{lem}
{\bf Proof:}
We have to show that there is no homoclinic orbit with core $\delta$ 
such that $(o+) <_u (\delta) <_u (o-)$ or $(e-) <_u (\delta) <_u 
(e+)$, where $e = s$ if $s$ has an even number of minus 
signs or $o = s$ if this number is odd.  If the initial string in 
$\delta$ differs from $s$ then $\delta$ can not be between the 
sequences $(s-)$ and $(s+)$, therefore $\delta = s\dots$. 
It is simple to see that 
\[ 
        \sdot e+\dots  \ge_u  \sdot e+-+^\infty, \quad 
        \sdot e-\dots  \le_u  \sdot e--+^\infty \; ,
\]
and similarly
\[ 
        \sdot o+\dots  \le_u  \sdot o+-+^\infty, \quad 
        \sdot o-\dots  \ge_u  \sdot o--+^\infty \; .
\]
Since $\delta = s\dots$ it must be of one of the forms on the left 
hand sides, but then the inequalities show that it is not between 
$(s+)$ and $(s-)$ hence they must be neighbors.  Conversely, suppose 
we have two neighboring homoclinic orbits (on $U$) $\sdot a$ and 
$\sdot b$ with $\sdot a <_u \sdot b$.  They must differ in at least 
one symbol so call the first such difference $x$.  Their leading 
common symbols are denoted by $s$, so that $a=s x\alpha$ and $b=s \bar 
x\beta$ for some sequences $\alpha$ and $\beta$, where $\bar x$ is the 
opposite symbol to $x$.  Applying the ordering relation to the 
possible combinations of $s$ and $x$ gives either
\[
  \sdot e - \alpha <_u \sdot (ey)-+^\infty <_u \sdot e + \beta \quad \mbox{or} \quad
  \sdot o + \alpha <_u \sdot (oy)-+^\infty <_u \sdot o- \beta \;,
\] 
where the choice of the symbol $y$ depends on whether $s$ is even or 
odd and whether $\alpha$ and $\beta$ are $-+^\infty$.  Specifically, 
choose $y=+$ if either $s=e$ and $\beta \ne -+^\infty$ or $s=o$ and 
$\alpha \ne -+^\infty$.  Choose $y=-$ if either $s=e$ and $\alpha \ne 
-+^\infty$ or $s=o$ and $\beta \ne -+^\infty$.  If neither $\alpha$ 
nor $\beta$ are $-+^\infty$ either choice for $y$ works.  When either 
$\alpha$, $\beta$, or both differ from $-+^\infty$ we have constructed 
an orbit
$ 
	\sdot (sy)-+^\infty  
$
which is between $\sdot a$ and $\sdot b$---hence 
$\sdot a$ and $\sdot b$ are {\it not} neighbors. But this is a contradiction
so $\alpha$, $\beta = -+^\infty$. \qed

For a bifurcation to occur it is not enough that the orbits be 
neighbors on $U$, but they must be double neighbors.  
In the reversible case this almost 
gives the proof of Conjecture~\ref{con:firstbif}, but here we are 
working in the smaller class of orbits homoclinic to $p$, the hyperbolic
fixed point. 
So far we did not make use of the reversibility of the map, i.e.,
the results are valid for all $b$. 
From this point on we will always only talk about the area preserving
case.
Note that the ordering is used in a range of parameters before the 
first bifurcation occurs, so the horseshoe ordering is still valid.
\begin{teo}
\label{thrm:firstbif}
In the area preserving \hen map
the first homoclinic bifurcation of the invariant manifolds of
        the fixed point $\per{+}$ is 
\[
        \bif{\sn}{\homoc{+}, \homoc{-}} \;.
\]
\end{teo}
{ \bf Proof:}
By Lemma~\ref{lem:neighbor} we know that all neighbors on $U$ in the complete 
horseshoe are of the form $(s\pm)$. For these sequences to be double neighbors
they must be neighbors on $S$ as well. 
By reversibility this is equivalent to the sequence and its reverse 
being neighbors on $U$.
But this only true if $s$ is empty. 
The only double neighbors in the complete horseshoe are therefore the two 
orbits $\homoc{+}$ and $\homoc{-}$. Therefore they must bifurcate first.
\qed

\begin{table}[bt] \centering
{ \begin{tabular}{c|c|c|c} 
Orbits            & $\ksn$ & $k_{\infty}$ & $\delta$  \\ 
\hline 
 $\per{-*-+^2}$    & 5.5517014388520 & 5.699160106302 &          \\
 $\per{-*-+^3}$    & 5.6793695105731 & 5.699306445540 & 7.45095  \\
 $\per{-*-+^4}$    & 5.6965039879058 & 5.699310669970 & 7.11409  \\
 $\per{-*-+^5}$    & 5.6989125149379 & 5.699310783741 & 7.04922  \\
 $\per{-*-+^6}$    & 5.6992541878224 & 5.699310786628 & 7.03706  \\
 $\per{-*-+^7}$    & 5.6993027411880 & 5.699310786699 & 7.03489  \\
 $\per{-*-+^8}$    & 5.6993096429803 & 5.699310786700 & 7.03452  \\
 $\per{-*-+^9}$    & 5.6993106241120 & 5.699310786700 & 7.03446  \\
 $\per{-*-+^{10}}$ & 5.6993107635871 & 
							& 7.03445  \\
 $\per{-*-+^{11}}$ & 5.6993107834145 & 
							& 7.03445  \\
\end{tabular}
}
\caption{Bifurcations in periodic approximations to the homoclinic type 
	1 orbit, which is the first orbit destroyed for $b=1$.  Here we use a 
	$*$ to denote both $+$ and $-$, giving both orbits involved in the 
	bifurcation.
          \label{tbl:exit} }
\end{table}

To approximate a homoclinic orbit, which possesses an infinite number 
of points in phase space by a periodic orbit with only a finite 
number of points we require that the Hausdorff distance of these two 
point sets vanishes as the period approaches infinity.
Thus for an orbit homoclinic to $\per{+}$, we study a sequence of 
approximating periodic orbits with an increasingly long string of $+$ 
symbols.  In  particular the rotational orbits 
given in \eq{rotorbits} converge to $\zeta$ and $\alpha$ in the limit.
  
In \Tblref{tbl:exit} we list the first 11 members of the sequence 
approximating the transit time $1$ homoclinic orbit, and the 
corresponding sequence of values, $\esn$ at which these orbits undergo 
a saddle-node bifurcation when $b = 1$.  These values converge 
geometrically to the parameter at which the homoclinic orbits 
bifurcate, and the ratio of successive differences ( a ``Feigenbaum 
ratio'') is computed in the fourth column of the table.  As is known 
theoretically for $b<1$ \cite{Curry82, Rob83} the convergence rate, 
$\delta$, approaches $\lambda$, the multiplier of the fixed point $p$.  
From our data, the convergence rate $\delta$ agrees up to 6 digits 
with the multiplier
\[
    \lambda \approx 7.0344478 
\]
of the fixed point $p$ when $k \approx 5.699310786700$.
Thus, our observations indicate that the convergence rate is given by 
the multiplier in the area preserving case as well, where to our 
knowledge no proof exists. 

The third column in the table is the 
extrapolation for the converged $k$ value, given by Aitken's 
$\Delta^{2}$ method
\[
    k_{\infty} = k_n - \frac{\Delta(k_{n})^2}{\Delta^2(k_{n})} \;,
\]
where $\Delta$ is the forward discrete difference operator.
Thus we see that there is a saddle-node bifurcation of the type $1$ homoclinic orbits, 
\[
	\bif{\sn} { +^\infty-(+)-+^\infty,  +^\infty-(-)-+^\infty } \;,
\]
at
\[
    \esn(1) \approx  0.418879233367 \quad \mbox{ or }\quad 
    \ksn(1) \approx  5.699310786700 \;.
\]
This also corresponds to the parameter value at which the topological 
horseshoe for the \hen map is destroyed, and is the value in 
\Fig{fig:horsebound} at $b=1$.

Since there is a sequence of saddle-node bifurcations that limit on 
the homoclinic bifurcation, there are elliptic islands arbitrarily 
close to the destruction of the horseshoe.  
This corresponds to an area preserving version
of the results of  Gavrilov and Silnikov \cite{Gav72,Gav73}
and Newhouse \cite{New74, New77}.

Since we can in principle follow every finite orbit from the 
anti-integrable limit we can begin to study the sequence of 
bifurcations that occur after the horseshoe is destroyed, see 
\Tblref{tbl:hobis}.  For example, the bifurcation diagram for all of 
the homoclinic orbits of type three or less is sketched in 
\Fig{fig:bifdiag}.  The vertical ordering in this sketch is the same 
as that on the segment $U$ with $\alpha$ and $\zeta$ shown.  The 
bifurcation diagram is highly influenced by the time-reversal symmetry 
of the area preserving \hen map---we will discuss this symmetry in the 
next section.  As expected from the general theory \cite{Rimmer78}, we 
observe three kinds of bifurcations:

\begin{description}
        \item[Symmetric saddle-node] bifurcations resulting in the creation
        of a 
        pair of type $t$ homoclinic orbits with opposite signatures. For 
        example, in \Fig{fig:bifdiag}, the type $3$ orbits with cores $(+++)$ 
        and $(-+-)$ are born in such a saddle-node at $k \approx 0.386$.

        \item[Pitchfork] bifurcations of type $t$ symmetric homoclinic orbits, 
        creating a pair of type $t$ asymmetric orbits that are related by 
        time reversal. For example, the $(-+-)$ orbit pitchforks at $k \approx 
        0.720$ creating the orbits $(-++)$ and $(++-)$. A pitchfork bifurcation
        requires triple neighbors to occur.
	The parent orbit of a homoclinic pitchfork bifurcations is 
	always created in a symmetric saddle-node bifurcation.
	Up to type 11 there are are only 9 symmetric saddle-node 
	bifurcations
	which do {\em not} undergo a homoclinic pitchfork bifurcation
	on their way to the AI limit.

        \item[Asymmetric saddle-node] bifurcations creating two symmetry 
	related 
        pairs of asymmetric orbits.  This bifurcation first occurs at 
        type $4$.  For example the two pairs $\{(-+--), (-+-+)\}$ and
        $\{(--+-),(+-+-)\}$ are created at $k \approx 5.18$. Generically,
        asymmetric saddle-node bifurcations require two pairs of double 
        neighbors to occur because of the symmetry.
\end{description}

The shaded region in \Fig{fig:bifdiag} represents the range of $k$ for 
which the area preserving \hen map exhibits a horseshoe.  Along the 
left edge we label each orbit with its core symbol sequence.

\Epsfig{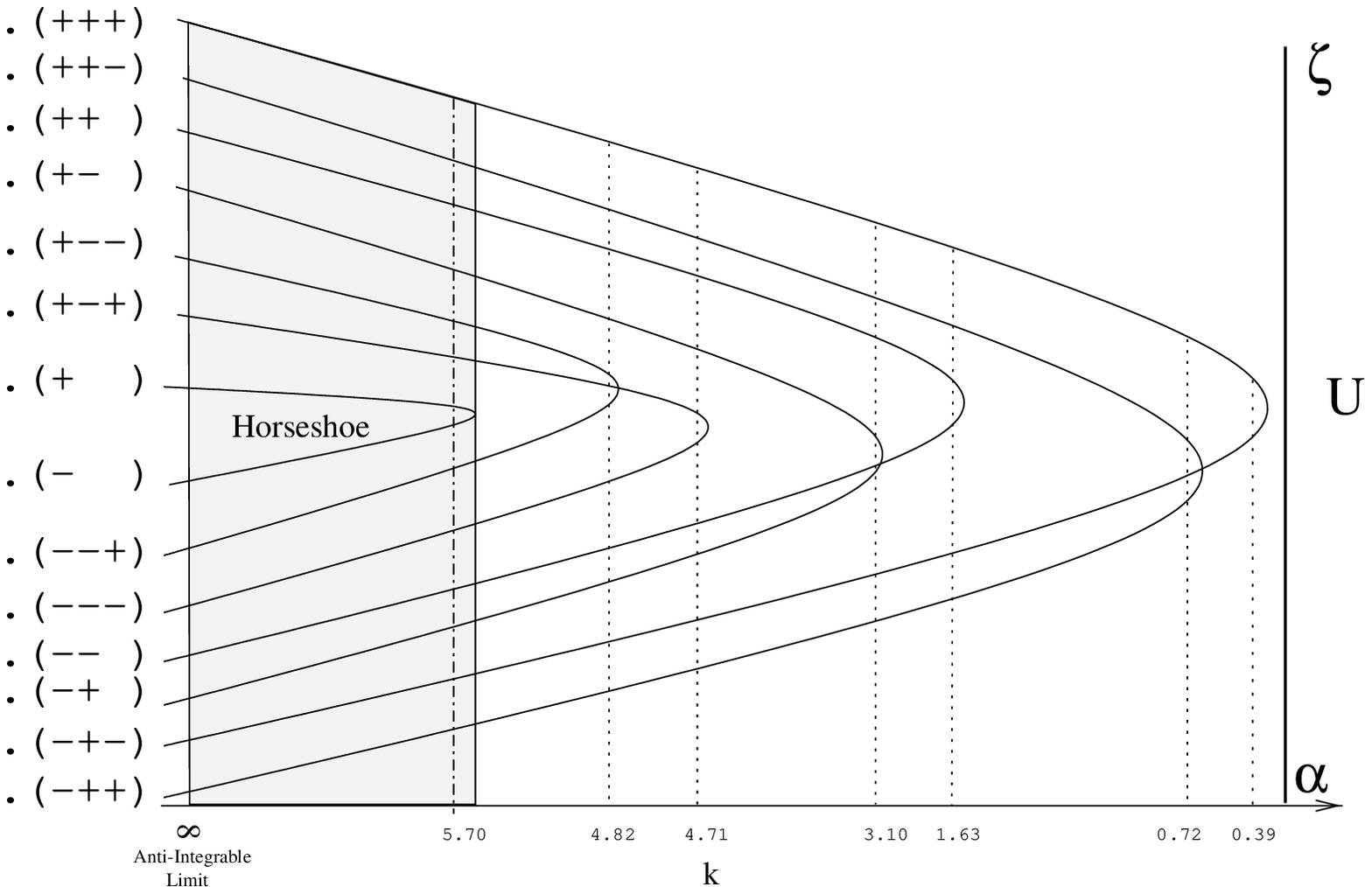}{tbp}
     {Sketch of bifurcations in the homoclinic orbits up to type 3 ($b=1$).} 
     {fig:bifdiag}{4in}

\begin{table}[btp]
      \centering
      \begin{tabular}{c|c|cc|c}
 Parent         & Type  & Child         & Child     & $k$-Value  \\  \hline 
                & $\sn$  & $(-++-)$     & $(++++)$  & -0.133474 \\
$(-++-)$        & $\pf$  & $(-+++)$     & $(+++-)$  & -0.044273 \\
                & $\sn$  & $(-+-)$      & $(+++)$   &  0.385556 \\
$(-+-)$         & $\pf$  & $(-++)$      & $(++-)$   &  0.719630 \\
                & $\sn$  & $(--)$       & $(++) $   &  1.627779 \\
$(--)$          & $\pf$  & $(-+)$       & $(+-) $   &  3.091505 \\
                & $\sn$  & $(+--+)$     & $(----)$  &  3.98213640 \\
$(----)$        & $\pf$  & $(---+)$     & $(+---)$  &  3.98213641 \\
                & $\sn$  & $(+-+)$      & $(---)$   &  4.706399 \\
$(+-+)$         & $\pf$  & $(--+)$      & $(+--)$   &  4.816792 \\
                &$\asn$  & $(-+-*)$ & $(*-+-)$  &  5.188561 \\
                &$\asn$  & $(*-++)$ & $(++-*)$  &  5.619922 \\
                & $\sn$  & $(+)$        & $(-)$     &  5.699311 
\end{tabular}
    \caption{Homoclinic bifurcations up to core length 4.}
\label{tbl:hobis}
\end{table}

The first type $t$ homoclinic orbits are created by a saddle-node 
bifurcation when the segment $f^{-t}(S)$ first intersects $U$. We 
denote this parameter value by $\ksn(t)$.  This marks the creation of 
the subset of the incoming lobe of the turnstile with transition time $t$ 
\cite{Meiss97}.  We observe that when $b=1$, this homoclinic 
saddle-node bifurcation is
\[
   \bif{\sn}{ (+^{t}) ,  (-+^{t-2}-)  } \quad \mbox{at\ }
		 \ksn(t) \;.
\]
Following this, the orbit $(-+^{t-2}-)$ undergoes a homoclinic pitchfork 
bifurcation at $\kpf(t)$, creating the pair
\[
  \ptcbif{ (-+^{t-2}-) }{\pf}{  (+^{t-1}-),  (-+^{t-1}) } \quad \mbox{at\ } \kpf(t) \;.
\]
However, when $b \ne 1$, the initial symmetric bifurcation and the 
following symmetry breaking pitchfork are replaced by a pair of 
nonsymmetric saddle-node bifurcations.  In this case the first type 
$t$ bifurcation is the homoclinic saddle-node
\[
    \bif{\sn} { (+^{t}),  (+^{t-1}-) } \;.
\]

According to the double neighbor lemma, certain bifurcations cannot 
occur prior to other homoclinic bifurcations because the corresponding
sequences block other sequences from being neighbors.
In order to determine which orbits are neighbors even 
beyond the first bifurcation we make
the assumption that the following symbolic ordering conjecture holds:
\begin{con}\label{con:orderpersists}
        The symbolic horseshoe ordering on the invariant manifolds 
        given in Lemma~\ref{orderAIL} persists.
\end{con}

The ordering relations 
give a unique construction of the order of the points on $U$ and $S$, 
and this implies that a schematic construction of the intersections 
of $f^{-t}(S)$ with $U$ can be constructed solely from a list of which 
orbits exist at a given parameter value. Such a schematic manifold 
plot is shown in  \Fig{fig:core5mf}, for all homoclinic orbits that 
exist at $k=5.53$ up to type $5$.

We can also construct a schematic bifurcation diagram for homoclinic 
orbits, as in \Fig{fig:core5bif}, by drawing a horizontal line from 
$k = \infty$ to the $k$-value at which a particular homoclinic orbit 
is destroyed---actually we stop the figure at $k=6$, since there are 
no bifurcations for larger $k$-values.  We order the homoclinic orbits 
vertically according to their unimodal ordering on $U$ as usual.  In 
this bifurcation diagram the vertical connections indicate which 
orbits eventually do become neighbors and bifurcate.  So as to avoid 
artificially crossing lines, we connect pairs of asymmetric saddle-nodes by 
lines at the right edge of the figure to indicate that they must 
bifurcate at the same $k$-value.

We say that a bifurcation {\it straddles} the centerline if
the pair of orbits involved are on either side of center of 
the $U$ ordering, or if one of the two pairs of an asymmetric 
saddle-node straddles the center line.

Through type $6$, each symmetric saddle node is followed by a 
pitchfork bifurcation, just as we observed in \Fig{fig:bifdiag}, with 
the exception of the very first bifurcation, $\bif{\sn}{(+),(-)}$, 
which corresponds to the smallest loop straddling the center in the 
figure. 
That this is in fact the smallest loop and therefore the first bifurcation
is the content of Theorem~\ref{thrm:firstbif}.

Moreover, it is remarkable, but perhaps misleading, that through type 
$6$ every bifurcation straddles the center.  Therefore all homoclinic 
bifurcations up to type 6 are forced by nesting around the center.  In 
particular this means that their unimodal ordering gives the 
bifurcation ordering, like in unimodal maps.

This simple forcing relation is destroyed with the 
appearance of a symmetric saddle-node without pitchfork of type $7$
(see \Tblref{tbl:gaps}.
Also at type $7$, there is an asymmetric saddle-node quadruple which 
does not straddle the center. Interestingly enough, this is the same 
bifurcation that marks the upper $k$ endpoint of one of the gaps that we 
discuss in \Secref{sec:entropy}.

\Epsfig{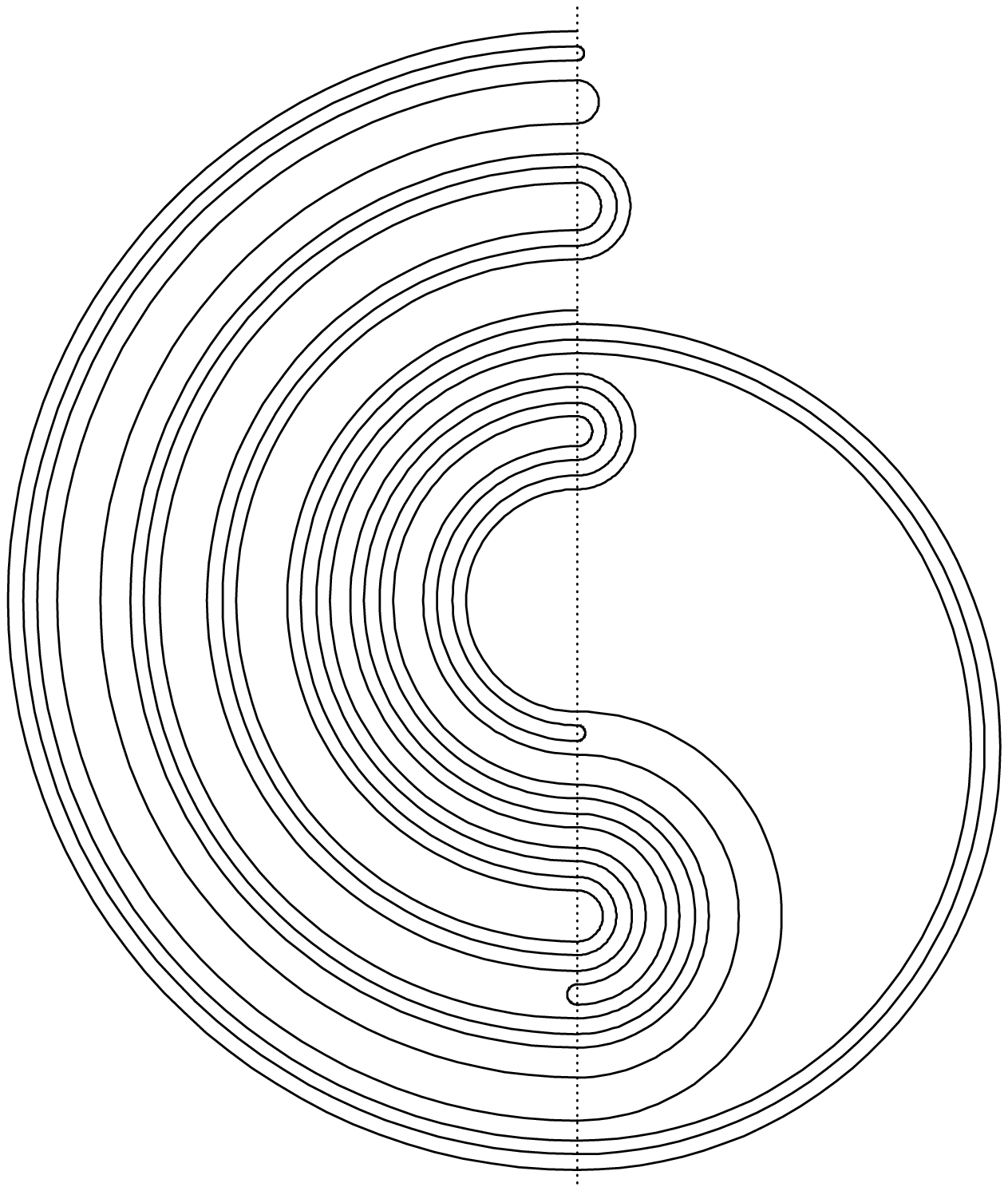}{p}
        {Schematic drawing of $U$ (dashed line) and $f^{-t}(S)$ (solid line) 
	up to type $5$ for $k=5.53$.}
        {fig:core5mf}{3in}

\Epsfig{core5bif.eps}{p}
        {Bifurcation diagram of homoclinic orbits up to type $5$ ($b=1$).
	 Types 1,2,3 are shown as dotted lines (recall Fig. 
	 \ref{fig:bifdiag}); type 4 is dashed; and type 5 is solid.}
        {fig:core5bif}{4in}

It is difficult to visualize the full homoclinic bifurcation diagram 
for larger type orbits.  To do so, we plot only the horizontal lines, 
to indicate the range of existence of an orbit; this diagram up to 
transition time 11 is given in \Fig{fig:core11}.  The approximate 
self-similarity in this picture seems to be related to some of the 
gaps we discuss in \Secref{sec:entropy}, namely those that are related to 
symmetric saddle-nodes without accompanying pitchforks of type 7, 9 
and 11.

\EpsfigR{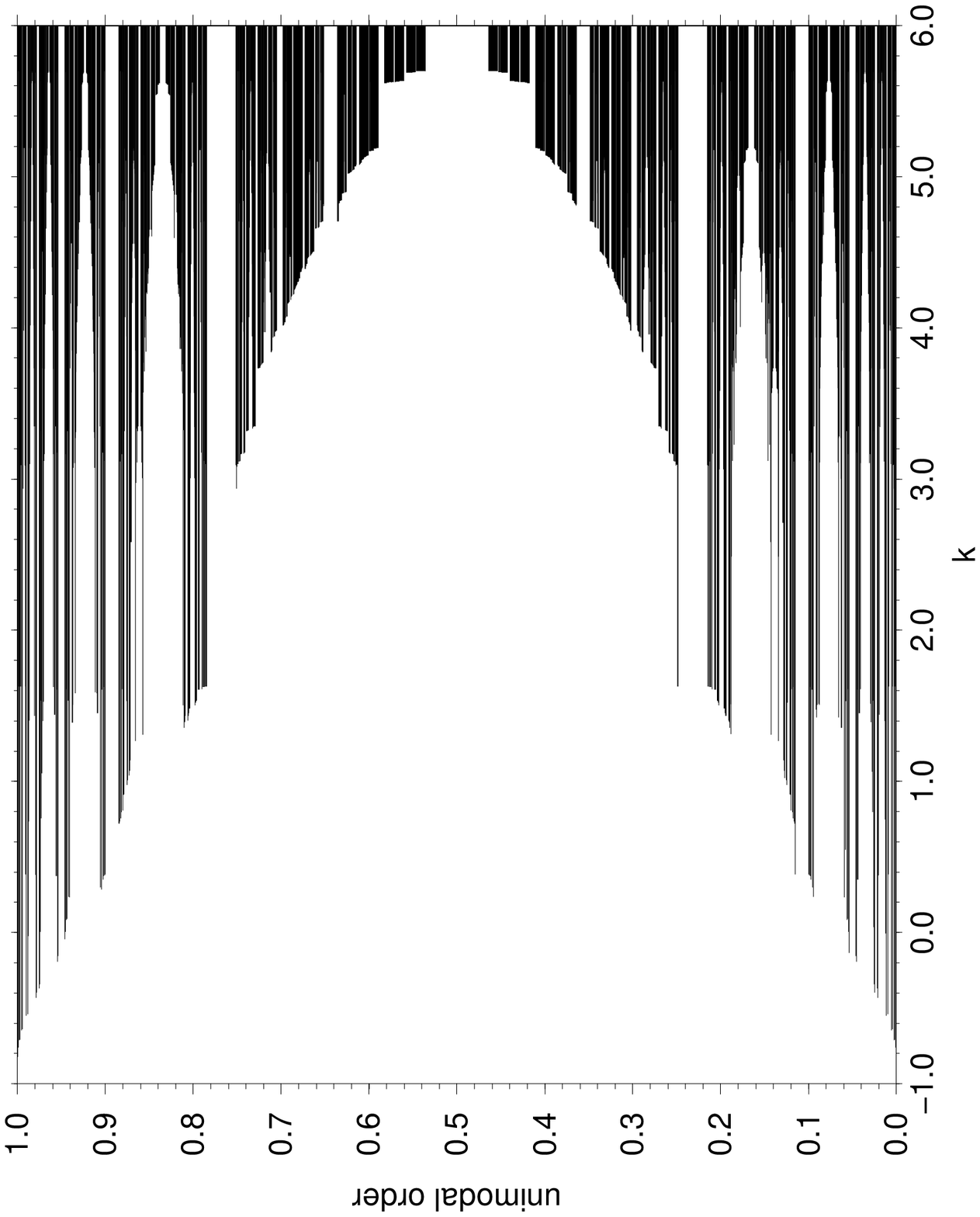}{bt}
   {Existence plot of homoclinic orbits up to type 11.  For each 
   homoclinic orbit a line is drawn from large $k$ to the parameter value 
   where this orbit is destroyed.  The vertical position of each line is 
   its formal position on $U$ according to the unimodal ordering.}
   {fig:core11}{6in}

\clearpage


\Sec{Symmetric homoclinic bifurcations}{sec:pitchfork}

As we mentioned above, the bifurcation diagram of the area preserving 
\hen map is restricted by the fact that the map has a time-reversal 
symmetry. Here we briefly recall a few well known facts about reversible 
maps \cite{Lamb98a}, and apply them to the study of homoclinic bifurcations.

A map $f$ has a time-reversal symmetry when it is diffeomorphic to
its inverse by:
\[ 
       Rf = f^{-1} R \;. 
\]
We call the map $R$ a reversor for $f$.  Often, as in our case,
the reversor 
is an involution, $R^{2}=I$.  Note that each of the maps $f^{t}R$ is 
also a reversor, in particular, we call $fR$ the complementary 
reversor to $R$.  The fixed set of a reversor
\[
     \fix(R) = \{ x : Rx = x \} \;,
\]
is of particular interest.  For the case when $R$ is an orientation 
reversing involution of the plane $\fix(R)$ is always a curve 
that goes through infinity, thus dividing the plane into two pieces 
\cite{MacKay93}.

A reversor maps an orbit $\ldots z_{t-1}, z_{t},z_{t+1}\ldots$ of the 
map onto another orbit $\ldots Rz_{t+1}, Rz_{t}, Rz_{t-1}\ldots$.  A 
symmetric orbit is defined as one that is mapped onto itself by $R$.  
It is easy to see that any symmetric orbit must have points on 
$\fix(R) \cup \fix(fR)$ and conversely. Moreover, if the orbit is not 
periodic, it has a unique point on one of these fixed sets, and if it 
is periodic it has exactly two points on the fixed sets \cite{Lamb98b}.

Reversible maps need not be area preserving, though the multipliers of 
an orbit and its symmetric partner must be reciprocals of one another.  
Application of this to the fixed points gives that the \hen map 
is reversible only when $b = \pm 1$.
For a symmetric orbit reversibility implies that
the product of the multipliers must be one.  
For the case $b=\pm 1$ a reversor is $R(x,y) = (-y,-x)$, and 
a complementary reversor $fR(x,y) = (-x-k+y^{2}, y)$.  The fixed 
curves are
\begin{eqnarray*}
        \fix(R)  &=& \{(x,y): x=-y\}  \;,\\
        \fix(fR) &=& \{(x,y): x = \frac12 (y^{2}-k) \} \;.
\end{eqnarray*}

Suppose that $p$ is a symmetric, hyperbolic fixed point of a 
reversible map.  Then, as pointed out by Devaney \cite{Devaney}, the 
stable and unstable manifolds of the map are related by $R$:

\begin{lem} \label{th:symman}
  Let $\Wu$ and $\Ws$ be the stable and unstable manifolds of a 
  symmetric fixed point $p$. Then $R\Wu(p,\balpha] = \Ws(p,R\balpha]$.
\end{lem}
{\bf Proof:} By definition, when $\balpha \in \Wu$, then 
$f^{-t}(\balpha) \rightarrow p$ as $t \rightarrow \infty$.  Then 
$Rf^{-t}(\balpha)= f^{t}(R\balpha) \rightarrow Rp = p$.  Thus, 
$R\balpha \in \Ws$. Since $R$ is a diffeomorphism, $R\Wu(p,\balpha] = 
\Ws(p,R\balpha]$. \qed

\begin{cor}
  If $\Wu$ intersects the fixed set of a reversor, then the intersection
  point is homoclinic.
\end{cor}
 
Homoclinic orbits of symmetric periodic orbits either come in 
symmetric pairs or are symmetric, 
and there must exist symmetric homoclinic orbits:

\begin{lem} 
  Let $p$ be a symmetric, hyperbolic fixed point, and $\balpha$ a  
  homoclinic point, and suppose that $R$ is an orientation reversing 
  involution. Then $R\balpha$ is also a homoclinic point.  
  Moreover, there exist symmetric homoclinic points on $\fix(R)$ and 
  $\fix(fR)$.
\end{lem}
{\bf Proof:} By Lemma \ref{th:symman}, since $\balpha \in \Ws \cap \Wu$ 
then $R\balpha \in \Wu \cap \Ws$, so it is homoclinic as well.  Since 
$\fix(R)$ divides the plane and $\balpha$ and $R\balpha$ are on opposite 
sides of this curve, the segment $\Wu[\balpha,R\balpha]$ must cross 
$\fix(R)$, and the crossing point is necessarily homoclinic and
symmetric.  We can argue similarly for $fR$. \qedd \\

As is well known, pitchfork bifurcations occur with codimension one in 
maps with a symmetry \cite{Rimmer78}.  This occurs for homoclinic 
bifurcations as well, as was suggested in \cite{Rom95}.  We observed 
such pitchfork bifurcations in \Fig{fig:bifdiag}.  A pitchfork 
typically occurs after a symmetric, type $t > 1$, saddle-node 
bifurcation creates a ``tip'' of $\Ws$ inside the entry lobe of the 
turnstile.  As this tip grows, one would normally expect it to bend 
around, as sketched in \Fig{fig:saddle-node}, creating more type $t$ 
homoclinic points by saddle-node bifurcation.  In fact, it is a simple 
consequence of the linear ordering along $\Wu$ and $\Ws$ combined
with reversibility that a single 
saddle-node bifurcation like that sketched in \Fig{fig:saddle-node} is 
impossible:

\Epsfig{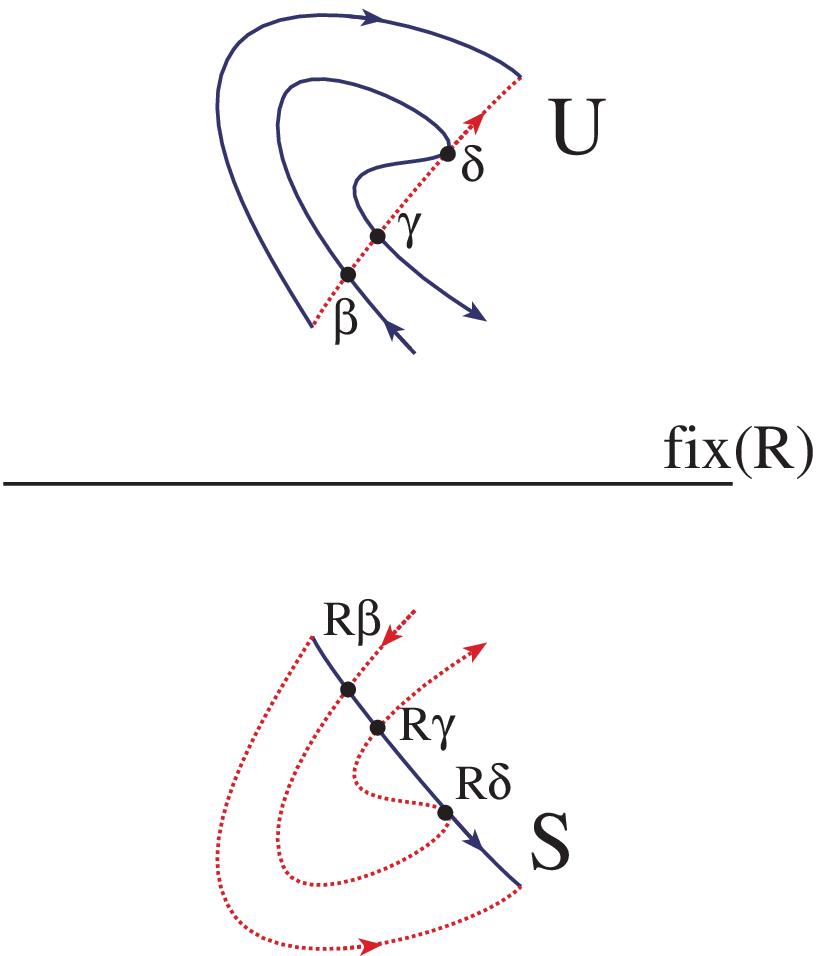}{bt}
        {Impossibility of the second symmetric bifurcation as described
        in \Th{pitchfork}.
        Stable manifolds are shown as solid and unstable 
        manifolds as dashed curves.  Reflection of a tangency at $\delta$ 
        gives a tangency at $R\delta$ that is ordered incorrectly.} 
        {fig:saddle-node}{3in}

\begin{teo} \label{pitchfork}
   Suppose that $f$ is an orientation preserving, reversible map, with a 
   symmetric fixed point $p$, and $S$ and $U = RS$ are segments of its 
   stable and unstable manifold bounded by adjacent primary homoclinic 
   orbits.  Suppose that a pair of symmetric homoclinic points on $U$, 
   $\beta <_{s} \gamma$ are created in a saddle-node bifurcation.  Then 
   it is impossible for there to be a single saddle-node bifurcation as a 
   tangency of $\Ws(\beta,\gamma)$ with either piece of $U \setminus 
   \Wu[\beta,\gamma]$.
\end{teo}

{\bf Proof:} Since $\beta <_{s} \gamma$, and $R\Ws = \Wu$, we have 
$R\beta <_{u} R\gamma$.  Suppose that $\beta$ and $\gamma$ have 
transition time $t$.  Then $f^{t}(\beta) \in S$, but since $\beta$ is 
symmetric this point must be the same as $R\beta$.  Thus 
$f^{-t}(R\beta) = \beta$, and similarly for $\gamma$.  Since the 
ordering is preserved by iteration, then $\beta <_{u}\gamma$.  Now 
suppose there is a tangency at a point $\delta = \Ws(\beta,\gamma) 
\cap (U \setminus \Wu[\beta,\gamma])$, i.e., not between $\beta$ and $\gamma$.  
We sketch such a configuration 
in \Fig{fig:saddle-node}.  Thus $\beta <_{s} \delta <_{s} \gamma$.  By 
symmetry, $R\beta <_{u} R\delta <_{u} R\gamma$.  Since the ordering is 
preserved by iteration, we have $\beta <_{u} f^{-t}(R\delta) <_{u} 
\gamma$.  Thus $f^{-t}(R\delta)\in \Wu(\beta,\gamma)$ and so this 
point is not $\delta$ (consequently the orbit of $\delta$ is not 
symmetric).  Since the manifolds are tangent at $\delta$, they are 
also tangent at $R\delta$ by symmetry.  Thus there is a second, 
simultaneous tangency, on $U$ at $f^{-t}(R\delta)$ which contradicts 
the assumption that a single tangency occurs.  
\qedd \\

There are three possible resolutions: first one of the two orbits, 
$\beta$ or $\gamma$ could undergo a pitchfork bifurcation creating a 
symmetry related pair of homoclinic orbits.  For example, 
\Fig{fig:pitchfork} shows part of the homoclinic tangle at a 
parameter value where the type two homoclinic orbit with core sequence 
$(-\,-)$ pitchforks.  As $k$ increases this results in the creation of 
a pair of type 2 orbits with cores $(-+)$ and $(+-)$, see 
\Fig{fig:pitch_plus}.  Note that the new orbits are not symmetric, but 
that the reversal of $(-+)$ is $(+-)$, so they form a symmetric pair.

\Epsfig{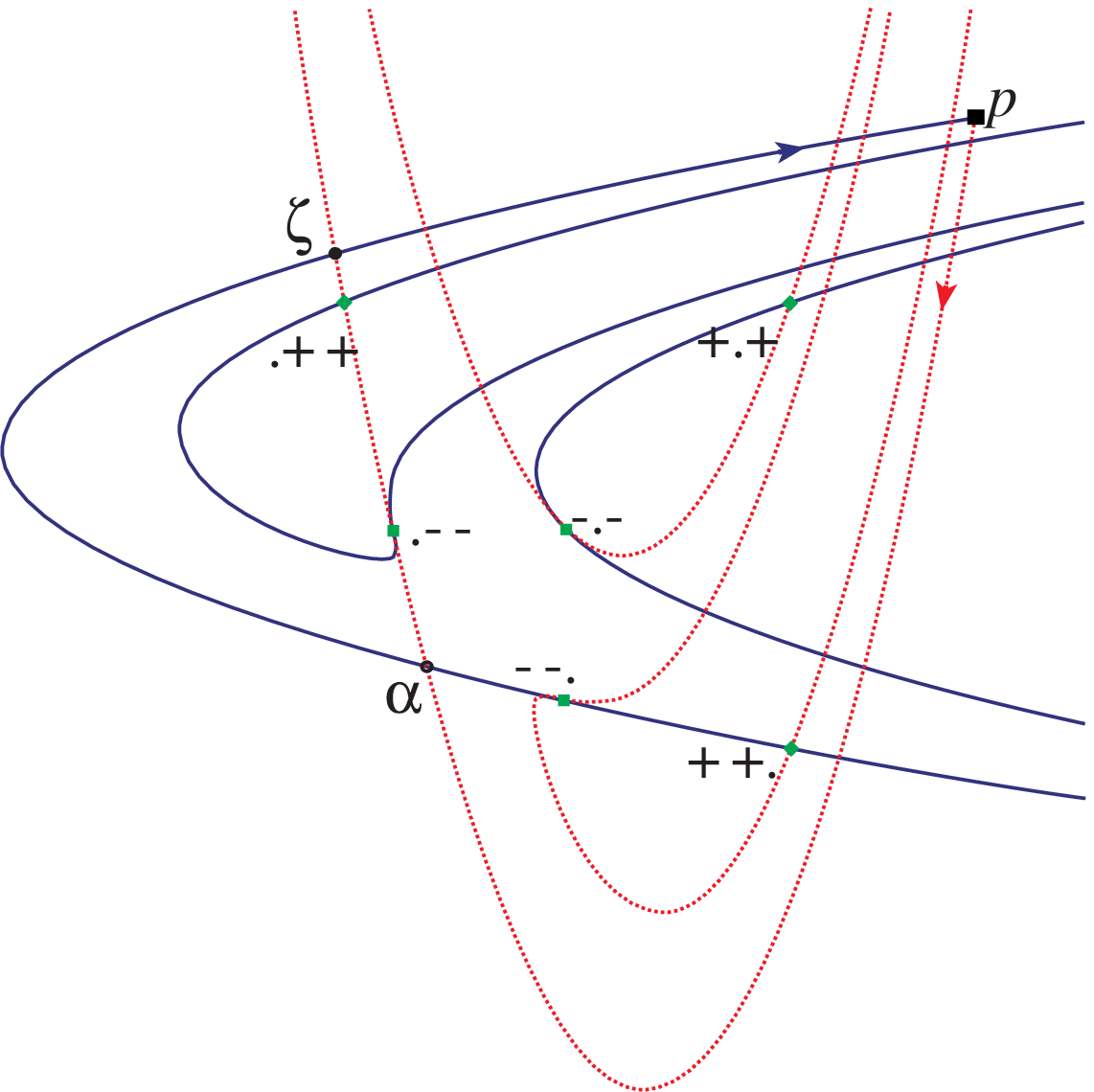}{tbp}
        {Stable and unstable manifolds for the $\per{+}$ fixed point of the 
        \hen map at $b=1$ and $k=3.09151$, where there is a cubic tangency of 
        the manifolds at the $+^{\infty}-(-\,-)-+^{\infty}$ homoclinic orbit.} 
        {fig:pitchfork}{3in}

\Epsfig{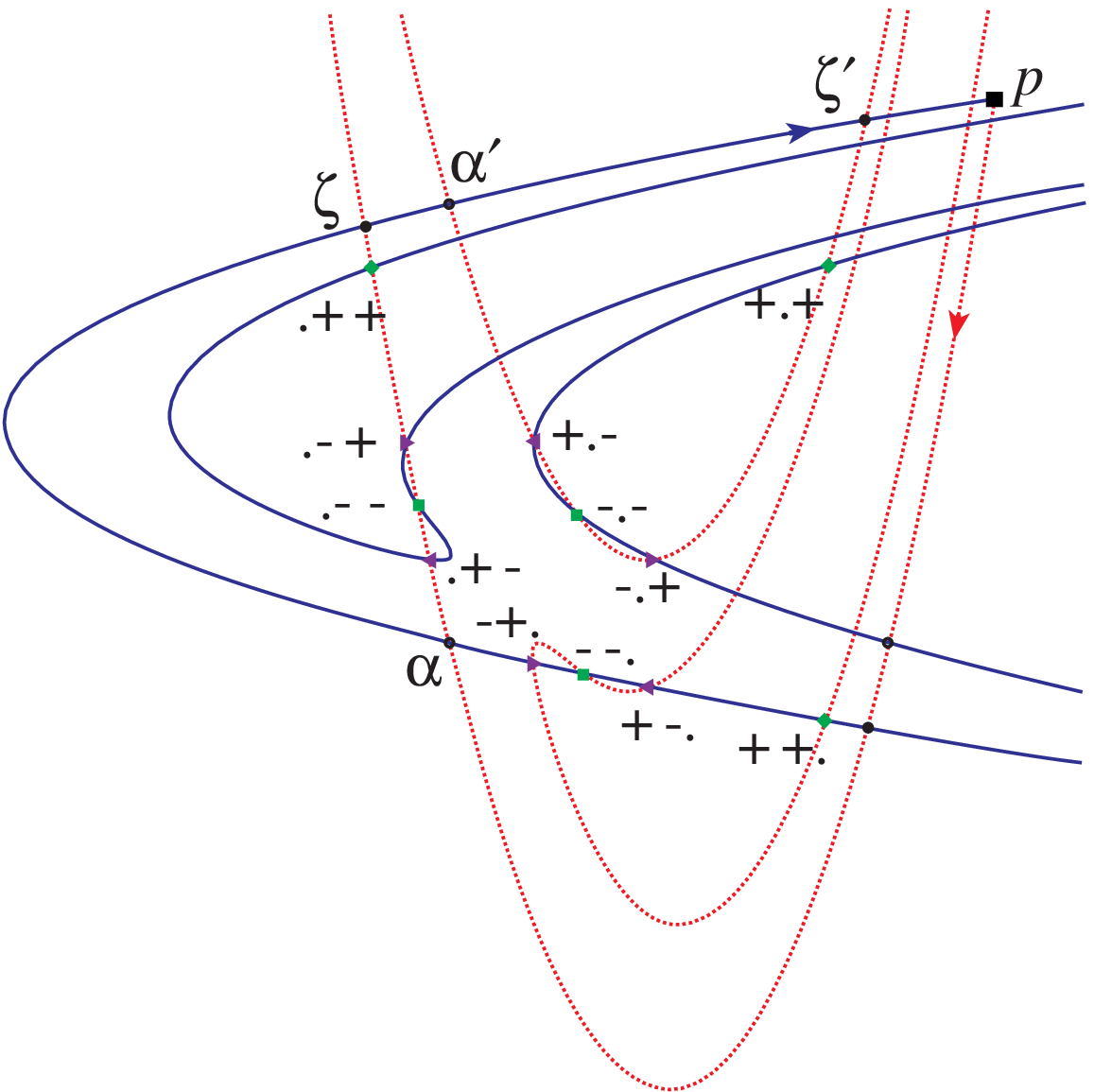}{tbp}
        {Type $2$ homoclinic orbits of the \hen Map at $k=3.5$.} 
        {fig:pitch_plus}{3in}
        
The second possible bifurcation is a single-saddle node on the 
segment $\Wu(\beta,\gamma)$; this happens, for example, whenever a 
``tip'' of an iterate of $S$ returns to $U$.  This first occurs at 
type $3$, for the bifurcation $\bif{\sn}{(*-*)}$.  We sketch a 
similar case, at type $4$, $\bif{\sn}{(*--*)}$, in \Fig{fig:asn}
which occurs at $k\approx 3.982$.

The third possible bifurcation is a pair of asymmetric saddle-node 
bifurcations.  This first occurs for homoclinic orbits of type $4$.  
For example, the bifurcations $\bif{\sn}{\homo{+-+-},\homo{--+-}}$ 
and its time-reverse, $\bif{\sn}{\homo{-+-+},\homo{-+--}}$ occur at 
$k\approx 5.1886$. We sketch $U$ and $f^{-4}(S)$ at this bifurcation in 
\Fig{fig:asn}. This bifurcation also corresponds to the lower 
endpoint of an apparently hyperbolic parameter interval for the \hen 
map, as we discuss in the next section.

Note that the antimonotonic bifurcations shown to exist in the area 
contracting  case \cite{Yorke92} are exactly forbidden by this theorem.

\Epsfig{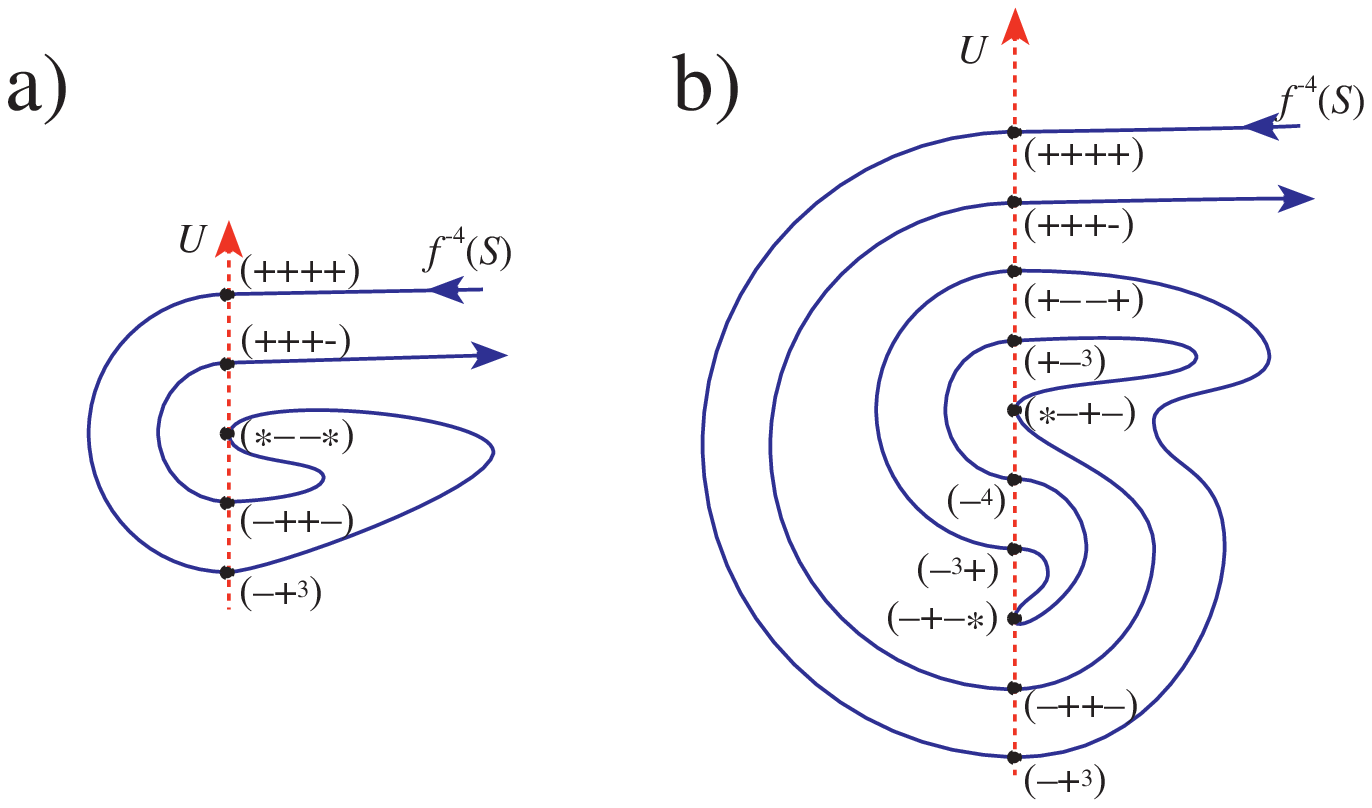}{bt}
        {Sketch of two possible homoclinic saddle-node bifurcations of type 
        four.  A symmetric saddle-node creating $(*--*)$ occurs on 
        $W^u((-++-),(++++))$ in (a).  An asymmetric saddle-node creating 
        $(*-+-)$ and $(-+-*)$ occurs with one point on $W^u((-^4),(+--+))$ in 
        (b).}
        {fig:asn}{6in}
        
A symmetric saddle-node followed by a pitchfork is a common 
bifurcation.  For example, the parameter values, $\ksn(t)$, at which 
the first type $t$ homoclinic orbit is created decrease monotonically with 
$t$.  Thus at $\ksn(t)$ the first type $t$ orbit is born and there are 
no homoclinic orbits with type less than $t$.  For $t>1$, at 
$\ksn(t-1)$ the segment $f^{t-1}(U)$ must intersect with $S$, so that 
$f^{t}(U)$ intersects with $f(S)$.  In order for this to happen (when 
$b=1$), there is a pitchfork bifurcation for $\kpf(t) \in [\ksn(t), 
\ksn(t-1)]$ of the type $t$ homoclinic orbit $\homoc{-+^{t-2}-}$ 
giving rise to the pair of orbits with symbol sequences
\[
     \ptcbif{\homo{-+^{t-2}-}}{\pf}{ \homo{-+^{t-1}} , \homo{+^{t-1}-}}  \;.
\]
We see that the children of this bifurcation differ from their parent 
in a single symbol and they differ from each other in two symbols.  
\Tblref{tbl:pitch} lists the first few such homoclinic bifurcation
 values obtained by extrapolation of the first few members of 
the approximating orbit sequence.

%
%
\begin{table}[tb]
\centering
{
  \begin{tabular}{c|c|l|c|l}   
   \tblrule $t$  & Core  &  $\ksn(t)$ & Pitchfork Children &  $\kpf(t)$ \\  \hline 
   \tblrule 1  & $(*)$     & \:5.69931078670  & & \\ 
   \tblrule 2  & $(**)$  & \:1.62777931098    &  $(-+),    (+-)$  & \:3.09150542113 \\ 
   \tblrule 3  & $(*+*) $  & \:0.38555621701  & $(-+^2),  (+^2-)$& \:0.71963023592 \\ 
   \tblrule 4  & $(*+^2*)$ &-0.13347378530  & $(-+^3),  (+^3-)$  &-0.04427324816 \\ 
   \tblrule 5  & $(*+^3*)$ &-0.39678970175  & $(-+^4),  (+^4-)$  &-0.36787481134 \\ 
   \tblrule 6  & $(*+^4*)$ &-0.54918558488  & $(-+^5),  (+^5-)$  &-0.53740149261 \\ 
   \tblrule 7  & $(*+^5*)$ &-0.64623270965  & $(-+^6),  (+^6-)$  &-0.64032496327 \\ 
   \tblrule 8  & $(*+^6*)$ &-0.71262572399  & $(-+^7),  (+^7-)$  &-0.70916824264\\ 
   \tblrule 9  & $(*+^7*)$ &-0.76055766670  & $(-+^8),  (+^8-)$  &-0.75830622014\\ 
   \tblrule 10 & $(*+^8*)$ &-0.79659407362  & $(-+^9),  (+^9-)$  &-0.79501732767\\ 
  \end{tabular}
} 
\caption{Pitchfork bifurcations from the first type $t$ orbits up to 
type 10.} 
         \label{tbl:pitch}
\end{table}

The distance (in $k$) between the birth of the type $t$ 
orbit and its pitchfork bifurcation shrinks to zero as the type increases.


\Sec{Intervals with no Bifurcations}{sec:entropy}

Davis, MacKay and Sannami (DMS) \cite{Davis91} used the numerical 
method of Biham and Wenzel \cite{Biham89} to compute the periodic 
orbits for the area preserving \hen map.  They showed that up to 
period $20$, there are intervals of parameter where there appear to be 
no orbits created or destroyed.  They studied a particular parameter 
interval near the destruction of the horseshoe, and elucidated the 
symbolic dynamics of the corresponding homoclinic tangle.  We will 
refer to this interval as the DMS gap.  Though the method of Biham and 
Wenzel is guaranteed to work close enough to the AI limit 
\cite{Sterling98a}, it can fail \cite{Grassberger89}.  We tested the 
DMS results using our continuation technique.  The use of parallel 
computation allowed us to extend the original experiment by an order 
of magnitude in size so that we followed all orbits up to period 
$24$---recall from \Tblref{tbl:periods} that there are a 
total of $1{,}465{,}020$ possible orbits.  We verify the DMS results and 
identify the symbol sequences of the orbits that form the boundaries 
of the DMS gap.

\Epsfig{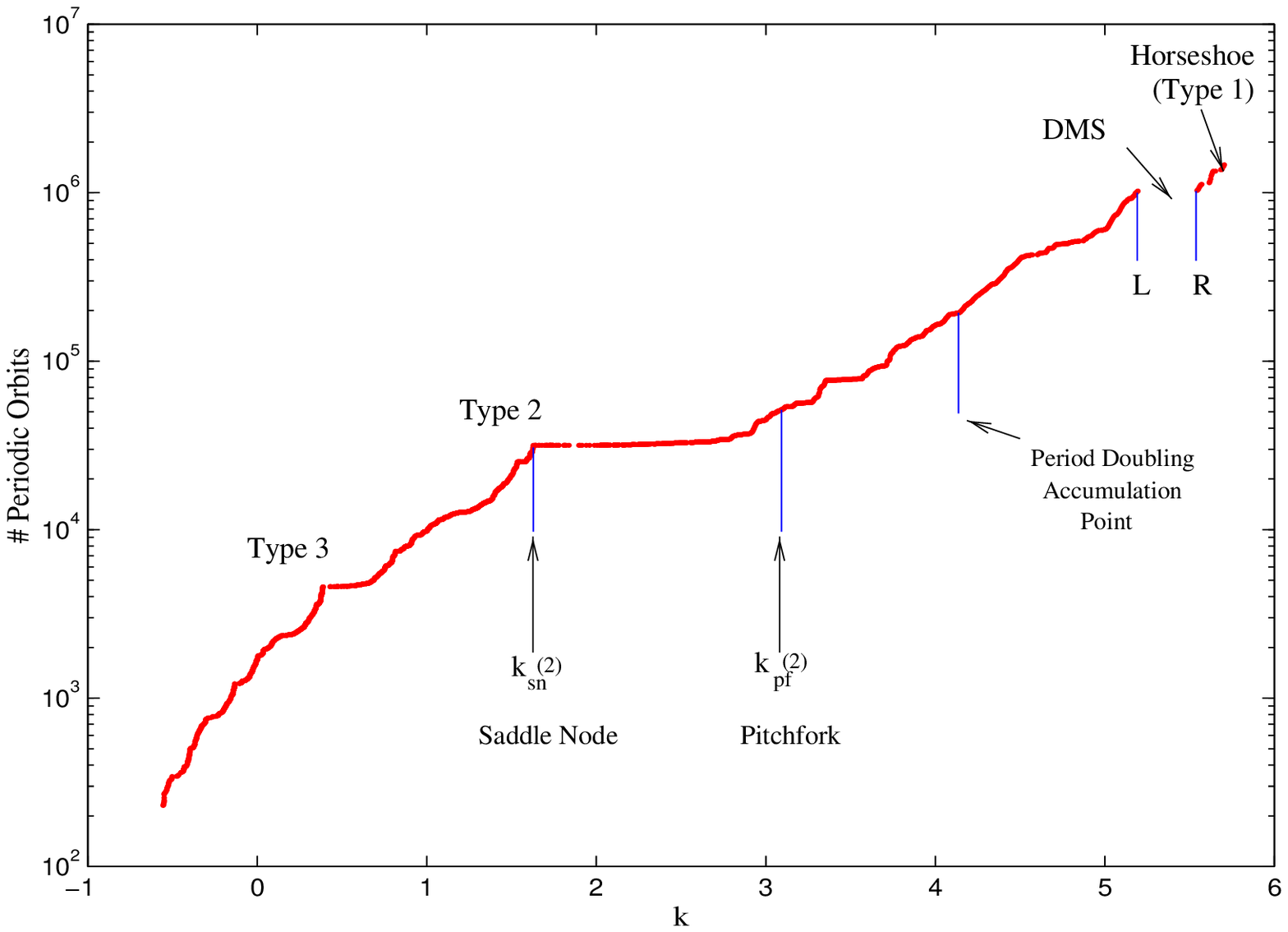}{tb}
      {Number of periodic orbits of the \hen map up to period 24, for $b=1$.
       The maximal number is reached at 
      $k\approx 5.69931078745$, when the horseshoe is formed. The endpoints of
      the largest gap studied by DMS are labeled by $L$ and $R$.}
      {fig:exitplateaus}{6in}

In our experiment we follow all orbits up to period 24 and record the 
minimal parameter values at which they are destroyed.  We then assume 
that each orbit exists only up to that value of $k$; this procedure is 
not entirely correct, because a few orbits loop back and forth in 
parameter under continuation.  This is related to the vanishing of 
twist in the neighborhood of a period tripling bifurcation 
\cite{DMS99}.  However, the number of low period orbits for which this 
happens is very small.  

In \Fig{fig:exitplateaus} we show the number 
of orbits that exist as a function of $k$, with the caveat that no 
value is plotted if the number of orbits does not change from the 
previously plotted point. This plot is equivalent to that of DMS, except 
that we leave gaps in the intervals where there are no bifurcations.

At the anti-integrable limit the map exhibits a horseshoe so all of 
the periodic orbits are present.  As we move away from the 
anti-integrable limit we see a decline in the number of periodic 
orbits as orbits collide and are destroyed.  Flat intervals 
in \Fig{fig:exitplateaus} represent intervals of parameter where very 
few bifurcations occur.  Gaps in the plot indicate intervals of 
parameter where there are {\em no} bifurcations.  The creation of the 
first type $t$ homoclinic orbits gives rise to flat intervals.  We 
observe that the left endpoint of each of the larger flat intervals 
for $k < 3$ corresponds to $\ksn(t)$ for the saddle-node bifurcation 
of the first type $t$ homoclinic orbits; these are marked in 
\Fig{fig:exitplateaus} and in the enlargement, 
\Fig{fig:plateaus_zoom1}.  Similarly, the parameter values $\kpf(t)$ 
are also marked; note that these pitchfork bifurcations are located 
well beyond the right endpoints of the flat intervals.  Each of the 
gaps in the flat intervals for $k < 3$ must eventually fill in if we go to high 
enough period because in this range of $k$ the area preserving \hen 
map has an elliptic fixed point. Recall that an $m/n$ bifurcation 
from the elliptic fixed point occurs at the parameter values $k_{m/n}$ 
given in \eq{Romega}, and these values are dense in the interval $-1 
\le k \le 3$.  Moreover, invariant circles bifurcate from the elliptic 
fixed point for each $k_{\omega}$ for sufficiently irrational 
$\omega$.
The same argument can be used up to the end of the period doubling
cascade of the fixed point at $k \approx 4.13616680392$, since each 
period doubling creates an elliptic orbit.

There are a number of distinct gaps in \Fig{fig:exitplateaus}; the 3 
larger gaps were studied by DMS, especially the largest one, near 
$k=5.5$ indicated by $L$ and $R$ in \Fig{fig:exitplateaus}.  DMS 
conjecture that the dynamics in each gap is hyperbolic, and 
consequently there are no bifurcations in a gap.  Our numerical 
evidence, which extends their study by an order of magnitude, supports 
this conjecture.  Upon examining the orbits that limit on the 
endpoints of the gap up to period $24$, we can extrapolate and find 
that each of the five largest gaps is bounded by a homoclinic 
bifurcation, see \Tblref{tbl:gaps}.  Thus we see that the gaps do 
not fill-in with orbits converging on the homoclinic bifurcations, but 
we cannot rule out that there are other, unrelated period orbits with 
period larger than $24$ that are created at parameter values in the 
middle of a gap.
\begin{table}[tbp]
        \centering
        \begin{tabular}{ll|ll}
     Left Endpoint Core  & $k_{L}$         & Right Endpoint Core    & $k_{R}$       \\ \hline
    $(-+-*-+-) $         & 4.55931896797   & $(++-*-*-++)$          & 4.59567964802 \\
    $(+^{3}-+-*)$        & 4.84317164217   & $(+^{3}-*-+-++-)$      & 4.86795762007 \\
    $(-+-*) $            & 5.18851121215   & $(++-*-++)$            & 5.53765692812 \\
    $(-++-*-++-)$        & 5.56490867348   & $(++-*-+^{3})$         & 5.60872105039 \\
    $(-++-*)$            & 5.63190980280   & $(+^{3}-*-+^{3})$      & 5.67769222229 \\
        \end{tabular}   
        \caption{Homoclinic bifurcations bounding the gaps 
	in \Fig{fig:exitplateaus}} \label{tbl:gaps}
\end{table}

We observe that there are two types of bifurcations bounding the gaps: 
symmetric and asymmetric saddle-node bifurcations.  The asymmetric 
saddle-nodes result in the creation of two pairs of homoclinic orbits, 
the one listed in the table, and its time-reverse.  Typically we 
observe that symmetric saddle node bifurcations in homoclinic orbits 
are followed by pitchfork bifurcations.  In fact we observe that among 
all of the homoclinic orbits through type $11$, there are only $9$ 
special saddle-node bifurcations which are not followed by a pitchfork 
bifurcation.  We believe that each of these special bifurcations 
corresponds to the endpoint of a gap.  For example, the first type 1 
bifurcation is not followed by a pitchfork, and it gives the left 
endpoint of the gap corresponding to the horseshoe.  The four gap 
endpoints in \Tblref{tbl:gaps} that correspond to symmetric saddle 
nodes are each of this special type.  The four remaining special pairs 
are each of type 11, and of these at least two bound gaps of widths 
$\Delta k \approx 6(10)^{-3}$.  With our resolution it is not possible 
to clearly identify the final two as gap boundaries.

In \Fig{fig:plateaus_zoom1} we show an enlargement of 
\Fig{fig:exitplateaus} but also include the data from the subshift, 
$\sigmaF$.  In the upper right corner of \Fig{fig:plateaus_zoom1} we 
see the tail end of the exit time $2$ plateau.  We also labeled the 
first large gap in $\sigmaF$ after the subshift is 
destroyed---this is the subshift analog of the DMS gap.

\Epsfig{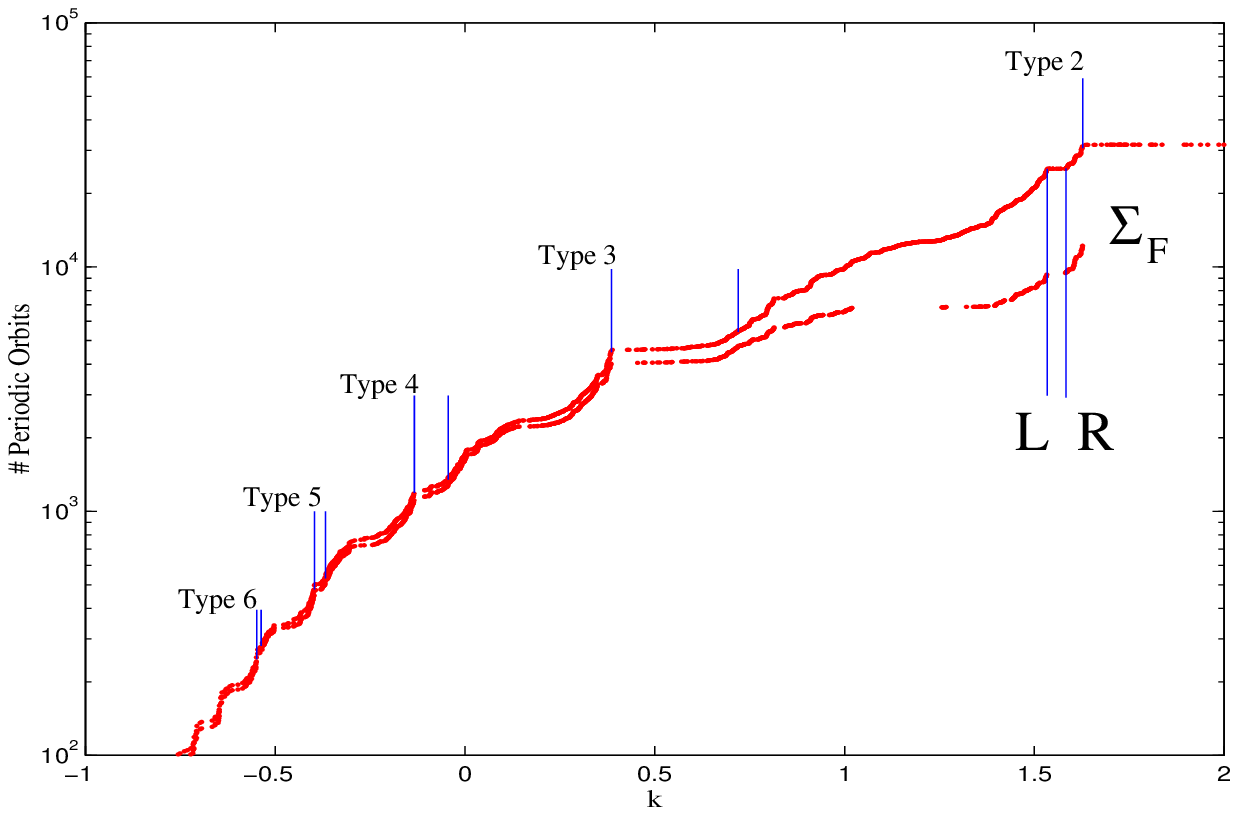}{ht}
      {Enlargement: number of periodic orbits up to period 24 for the 
      full shift and the subshift $\sigmaF$. The 
      left and right ends of each plateau correspond to $\ksn(t)$, and 
      $\kpf(t)$.}
      {fig:plateaus_zoom1}{6in} 

As in the DMS gap, the left and right boundaries, denoted $L$ and $R$,
correspond to a pair of homoclinic saddle-node bifurcations with the 
core sequences
\begin{eqnarray*} \label{subgap}
\bif{\asn}{ \homo{*+-++-}, \homo{-++-+*} } & \mbox{ at\ } &
	\ksn(L) \approx 1.533898312
\;, \\
\bif{\sn}{ \homo{+^3-++-+^3}, \homo{+^3-^4+^3} } & \mbox{ at\ } &
        \ksn(R) \approx   1.583387630 
\;.
\end{eqnarray*}
Note that the right endpoint of the gap corresponds to an 
orbit whose partner is not in the subshift!

The curves for all orbits and for the subshift are remarkably similar
and it appears that the growth of orbits in the subshift gives an accurate
representation of the overall growth of orbits in the full shift for this 
range of parameters. This is especially remarkable given that when all
the orbits exist, the subshift contains less than $1\%$ of the orbits in 
the full shift up to period 24.
The figure shows that for small $k$, the number of orbits in the full
shift is nearly a constant multiple of that in the subshift. 

Observing that the gaps are bounded by homoclinic orbits, 
we regenerated the orbit growth plot using {\it only} homoclinic orbits.
As expected, the gap structure and overall shape of \Fig{fig:exitplateaus} 
is almost completely captured by the homoclinic orbits alone.


\Sec{Conclusions}{sec:conc}

Continuation from an anti-integrable limit is an effective technique 
for studying orbits providing that there are no isolated bubbles in 
the bifurcation diagram.  
In \cite{Sterling98a} we applied the anti-integrable
theory to the \hen map to obtain a new proof of the well-known analytical
bound of Devaney and Nitecki \cite{Devaney79}. In \Th{thrm:seq} we apply
a similar argument to a restricted set of orbits to find an analytical
bound for the existence of a subshift of finite type. We present both
analytical bounds together with the optimal bounds generated 
numerically using our continuation method. We observe that the 
horseshoe is destroyed by a type one bifurcation that is homoclinic
in the orientation preserving case, and heteroclinic otherwise. In either case 
we conjecture that this bifurcation is the first bifurcation among {\it all} 
orbits of the \hen map as we recede from the anti-integrable limit.

With our continuation method, we are able to assign a ``global code'' 
to each orbit, by fixing the designation to that at the AI limit. In
the \hen map, we demonstrate that this AI code is equivalent to
the standard horseshoe code (when it exists), 
but it also gives a consistent way of assigning symbols to orbits 
{\it beyond} the destruction of the complete horseshoe. Remarkably, there
 appears to be a relationship between the AI codes for a number
of systems including billiards, twist maps of the cylinder and 
the \hen map. We will explore this relationship in a forthcoming 
paper \cite{Sterling98b}.

We relate the properties {\it transition time}, {\it type}, and {\it 
signature} of homoclinic orbits to properties of the core sequence.  
We also demonstrate that the ordering of the homoclinic orbits on the 
manifold segments $U$ and $S$ is the standard unimodal ordering.  The 
notion of {\it double neighbors} and lemmas \ref{lem:doubleN} and 
\ref{lem:ttime} give a necessary condition for a pair of homoclinic 
orbits to bifurcate.  Surprisingly, these also give a forcing relation 
that tells us which homoclinic bifurcations have to occur before other 
ones.  Showing that homoclinic bifurcations can only take place 
between {\it double neighbors} with the same core length, 
\Lem{lem:neighbor} gives a symbolic criterion for a pair of homoclinic 
orbits to be neighbors.  The ordering is certainly valid until the 
complete horseshoe is destroyed, which leads to the theorem that the 
first homoclinic bifurcation of the hyperbolic fixed point in the area 
preserving \hen map occurs between the pair of type one orbits.
When $b= \pm 1$, the \hen map has a symmetry and we discuss the 
mechanism by which pitchfork and asymmetric saddle node bifurcations 
occur.  The key ingredients to \Th{pitchfork} are the ordering on the 
manifolds and the existence of a reversor for the map.  As a result, 
the scenario of a tip of the manifold just repeatedly piercing through 
the other manifold (which is most natural in the case without 
symmetry) is impossible.  Among the possible alternatives are the 
occurrence of a pitchfork bifurcation or the creation of an asymmetric 
saddle-node pair, which are both not generic in the non-reversible 
case.

With our continuation technique we compute numerical 
values for various bifurcations of the homoclinic orbits up to type
eleven. We sketch the bifurcation diagram at type three, and
then use a simple algorithm to construct the much
more complex figures for higher core length. 

In contrast to our method for finding periodic orbits,
the Biham and Wenzel method \cite{Biham90,Biham91} 
is known to fail in certain cases \cite{Grassberger89}, and can only 
be justified in the neighborhood of the AI limit \cite{Sterling98a}.
Nevertheless, for the area preserving \hen map, we observe precisely
the same number of orbits in the main DMS gap 
using our technique as was reported by DMS using the 
Biham-Wenzel method \cite{Davis91}. We extend the original experiment
of DMS in two ways. First, we study an order of magnitude more orbits than
the original experiment and yet the gaps originally reported by DMS persist.
Second, we observe that homoclinic bifurcations are responsible for these gaps
and we list the symbolic labels of the orbits that form the gap endpoints in
Table \ref{tbl:gaps}. These gaps correspond to the creation and destruction
of parameter intervals where the dynamics of the area preserving \hen map 
appears to be conjugate to a subshift of finite type. 
We find a similar gap structure for a particular subshift, $\sigmaF$, and
list the symbolic labels for the homoclinic orbits that form the endpoints
of the gap that is the analog of the DMS gap. 
The role of the special symmetric saddle-node bifurcations without
an accompanying pitchfork bifurcation in this scenario remains
to be elucidated.

Many of our analytical results could be transferred from the area
preserving case $b=1$ to the orientation reversing 
case $b=-1$. In this case there also
exists a reversor, so that the theorem about the first bifurcation
and about impossibility of pitchfork bifurcations could be generalized
to this case. The main difference is that now we are not studying 
homoclinic orbits of a fixed point which is invariant under the
reversor, but instead heteroclinic orbits connecting fixed points
that are mapped into each other by the reversor. Correspondingly
the action of the reversor on symbol sequences is not just
reading them backwards, but it is reading backwards and flipping
each symbol.

Let us finally remark that we can extend the antimonotonicity 
result \cite{Yorke92} to the non-dissipative case $b=\pm 1$ 
because there are bifurcations that do occur in the ``wrong'' 
direction, i.e., orbits that are created when $k$ is decreased.
This is described in detail in a separate paper \cite{DMS99}
where it is related to the fact that in the neighborhood
of the period tripling bifurcation the twist generically vanishes.

%
\clearpage
\bibliographystyle{unsrt}  
\bibliography{Hen}

\end{document}